\documentclass{aa}
\usepackage{natbib}
\bibpunct{(}{)}{;}{a}{}{,}
\usepackage{lastpage} 
\usepackage{color}
\usepackage{xcolor}
\usepackage{float} 
\usepackage{graphicx}
\usepackage{hyperref}
\usepackage[normalem]{ulem}
\usepackage{xurl}  
\hypersetup{colorlinks=true,citecolor=blue,linkcolor=blue,anchorcolor=blue}

\usepackage{txfonts}

\begin{document}


    \title{The Complex Star-Forming Disk of M\,83: Mass Assembly and Chemical Evolution} 


   \author{Xiaoyu~Kang
          \inst{1}\fnmsep\thanks{kxyysl@ynao.ac.cn}
          \and
          Rolf-Peter~Kudritzki\inst{2,3}
          \and 
          Eva~Sextl\inst{2,4}
          \and
          Fenghui~Zhang\inst{1}
          \and
          Xiejin~Li\inst{1,5}
          \and 
          Ruixiang~Chang\inst{6,7}
          \and 
          Yunkun~Han\inst{1}
          }

   \institute{International Center of Supernovae (ICESUN), Yunnan Key Laboratory of Supernova Research, Yunnan Observatories, Chinese Academy of Sciences (CAS), Kunming 650216, China
         \and University Observatory Munich,
         Ludwig-Maximilian-Universit$\rm \ddot{a}$t M$\rm \ddot{u}$nchen, Scheinerstr. 1, 81679 Munich, Germany
         \and Institute for Astronomy, University of Hawaii, 2680 Woodlawn Drive, Honolulu, HI96822, USA
         \and Excellence Cluster ORIGINS, Boltzmannstr. 2, 85748 Garching, Germany
         \and University of Chinese Academy of Sciences, Beijing 100049, People’s Republic of China
         \and Shanghai Astronomical Observatory, Chinese Academy of Sciences, 80 Nandan Rd., Shanghai, 200030, People’s Republic of China
         \and Key Lab for Astrophysics, Shanghai, 200034, People’s Republic of China             
             }

   \date{Received 21 July 2026; accepted 26 November 2026}

 \abstract
{ 
The nearby face-on spiral galaxy M\,83 (NGC\,5236) shows a surprisingly complex radial metallicity distribution across its disk. To investigate the physical processes driving this unusual chemical enrichment, we construct a simple chemical evolution model tailored to M\,83. The model assumes that the disk forms progressively via accretion of primordial gas, incorporates a star formation law correlated to molecular gas, and includes metal-enriched outflows, with the gas infall timescale $\tau$ as the only free parameter.
By systematically exploring the parameter space and performing a $\chi^{2}$ minimization between model predictions and the observed radial profiles, we identify a viable model with a best-fitting infall timescale of $\tau = 2.7\,\rm{Gyr}$.
This model successfully reproduces the observed radial profiles of HI surface density, H$_2$ surface density and star formation rate (SFR) surface density across the entire disk, while the metallicity profiles of the ISM and young stars are matched only within the inner disk ($R/R_{25} \lesssim 1.2$). The model predictions for the present-day global quantities, including $M_{\rm{H}_2} = 3.67\times 10^9\,M_\odot$, $M_{\rm{HI}} = 6.83\times 10^9\,M_\odot$, and SFR $= 1.93\,M_\odot\,\rm{yr}^{-1}$, are in excellent agreement with observations and lie perfectly on the star-forming main sequence and the mass--metallicity relation. Furthermore, the model naturally captures an inside-out mass assembly history and predicts a $\sim 0.3$\,dex metallicity offset between young and total stellar populations, consistent with recent population synthesis studies.
However, the model fails to match the observed high metallicity in the outer disk ($R/R_{25} > 1.2$). To resolve this discrepancy, we test a ``toy'' model that introduces the accretion of pre-enriched gas ($Z_{\rm infall}/Z_\odot = 0.43$, as constrained by a systematic $\chi^{2}$ exploration) onto the outer disk, which successfully explains the extended, flat metal-rich HI disk and suggests that the infall of pre-enriched gas is the dominant enrichment mechanism in the outer regions of M\,83.
}




   \maketitle
%

\section{Introduction}
\label{sec:intro}

The disk components of spiral galaxies contain crucial information about their formation and evolutionary history. A particularly valuable nearby target is M\,83 (NGC\,5236), a nearly face-on spiral galaxy with a prominent bar \citep{Kormendy2004}, an inclination angle of $24^\circ$, and a position angle of $45^\circ \pm 5^\circ$ \citep{Comte1981}. Its distance has been measured as $d = 5.01 \pm 0.25$ Mpc from blue supergiants \citep{Bresolin2025}, and its isophotal radius is $6.44 \pm 0.05$ arcmin (corresponding to 9.38\,kpc at the adopted distance) \citep{deVaucouleurs1991}, making it one of the best nearby targets for studying the evolution and assembly history of massive spiral galaxies. 

The abundant multi-wavelength data available for M\,83, including ultraviolet/optical imaging (GALEX and HST), CO line emission (ALMA), 21\,cm radio mapping (THINGS/VLA), and integral-field spectroscopy (IFU, e.g. TYPHOON and MUSE), provide a comprehensive view of its interstellar medium (ISM) and stellar populations \citep[e.g.][]{Thilker2005, Walter2008, Blair2014, Leroy2021b, Grasha2022, Bruna2022, Long2022, Koda2023, Sextl2025, Sextl2026}. 
In particular, M\,83 hosts prominent extended HI and ultraviolet (XUV) disks \citep{Thilker2005, Koribalski2018, Wang2024}.
Leveraging these diverse datasets, we can simultaneously measure key physical quantities, including the radial distributions of surface densities of atomic hydrogen gas (HI), molecular hydrogen gas (H$_2$), star formation rate (SFR), as well as the metallicities of the ISM and young and old stars. These properties are essential for constraining galaxy evolution models.




The galaxy radial metallicity gradient is often regarded as the ``Rosetta Stone'' for probing the formation and evolutionary history of galaxies \citep{PB2000, Kudritzki2015, Belfiore2019, Kang2025}, because metallicity is determined by the extremely complex interplay between gas flows, star formation and stellar evolution. The results of multiple independent studies, including the detailed optical spectroscopy of blue supergiant stars \citep[BSGs;][]{Bresolin2016}, study of UV and optical integrated spectra of young massive clusters \citep[YMCs;][]{Hernandez2018, Hernandez2019}, a near-infrared investigation of red supergiant-dominated super star clusters \citep[SSCs;][]{Davies2017}, the analysis of metal emission and absorption features in the ISM \citep{Bresolin2005, Bresolin2016, Hernandez2021, Grasha2022}, and the population synthesis analysis with full-spectral fitting of TYPHOON IFU spectral data by \citet{Sextl2025}, reveal that the radial metallicity  distribution of M\,83 is extremely complex. This is illustrated by Figures \ref{fig:model_result} (lower right) and \ref{fig:Zfit_large}. The very central region shows areas of low metallicity, which are very likely the result of most recent metal poor gas infall and (see \citealt{Sextl2025} and \citealt{Sextl2026} for details) or the effect of of a recent dwarf galaxy-like interloper \citep{Diaz2006}. Beyond that the metallicity distribution of the optical disk of M\,83 is surprisingly flat with almost no gradient apparent from absorption and emission line studies of the ISM as well as the analysis of the tracers of the young stellar population.
Near the optical disk edge (at $R/R_{25} \approx 1.0$), there is a discontinuous jump, or ``break'' in the ISM metallicity. Beyond the ``break'', the ISM metallicity exhibits an approximately flat distribution across the outer disk, with the metallicity remaining at $\sim 1/3\,Z_{\odot}$ \citep{Bresolin2009}. 
These observed metallicity distribution features not only contradict the simple ``inside-out'' galaxy formation model \citep{Matteucci1989, Molla2017} but also strongly imply that the physical processes governing star formation and chemical enrichment undergo a fundamental alteration across the inner and outer sides of $R_{25}$ \citep{Bresolin2009, Bresolin2016, Kudritzki2014, Bell2026, Ogami2026, Kewley2026}.

To account for its extended HI disk and, in particular, its unusual radial metallicity distribution \citep{Bresolin2009,Heald2016}, several previous studies have explored various mechanisms. In their analysis of M\,83's ISM metallicity distribution, \citet{Bresolin2016} employed the chemical evolution model by \citet{Kudritzki2015}, where the metallicity is a function of the stellar-to-gas mass ratio. Most importantly, for the outer disk, they introduced an infall component with non-zero metallicity, representing material ejected from the inner disk that rains back onto the outer regions. This enables their model to successfully reproduce the observed ISM metallicity distribution.
\citet{Wang2022} attempted to apply their modified accretion disk model to M\,83, but found that it failed to reproduce the observed metallicity profile. The failure stems from the abrupt ``break'' in M\,83's ISM metallicity at a specific radius, which their analytical solution (a simple exponential plus a constant) is intrinsically unable to capture. 
Using TNG-50 simulations, \citet{Bell2026} constrained M\,83's merger history and explored the origin of its extended HI disk. Their results rule out recent accretion events as the primary gas source, instead favoring secular evolution or intergalactic medium (IGM) accretion as more plausible explanations. 
\citet{Ogami2026} utilized Subaru/HSC data to confirm a giant stellar stream (S-stream) in the southern part of M\,83, verifying its origin from a low-mass satellite galaxy merger. This minor merger event disturbed and redistributed the HI gas in M\,83's outer disk, triggering the formation of the XUV disk and explaining the anomalously high metallicity and flat gradient of the outer disk.

Metals are generated by stellar nucleosynthesis and act a fossil record of the star formation history (SFH) and gas flows within a galaxy. 
Galactic chemical evolution models, which adopt parameterized descriptions linking metallicity to the SFR and gas flow rates, have been widely used to reproduce the observed local and global chemical abundances, cold gas content, and SFR of the Milky Way and nearby disk galaxies \citep[e.g.][]{Yin2009, Marcon-Uchida2010, Chang2010, Kubryk2015, Matteucci2021, Spitoni2021, Kang2023, Kang2025}. The success of such model lies not only in fitting the observed data, but more importantly in their ability to constrain the fundamental physical processes that are difficult to measure directly.

To connect these general models to the specific case of M\,83, we construct a simple and computationally efficient galactic chemical evolution model tailored to this galaxy. To robustly reconcile its observed properties with its complex assembly history, this model must simultaneously reproduce multiple radial constraints—namely the surface densities of HI and H$_2$, the SFR surface density, and the metallicity profiles of both the ISM and young stars. Taking advantage of its high computational efficiency, we then systematically explore the parameter space to reconstruct the evolutionary histories of HI, H$_2$, SFR, and specific star formation rate (sSFR) in M\,83. 
Ultimately, this allows us to address key questions regarding its radial metallicity distribution and the dominant enrichment mechanisms operating in its disk, based on the reconstructed histories of its gas and star formation.




This paper is structured as follows. Section\,\ref{sec:model} presents the main ingredients of the chemical evolution model used in this work. Section\,\ref{sec:obv} presents the main observed data of M\,83 used to constrain the model. Section\,\ref{sec:result_dis} presents our results and  discussion. Section\,\ref{sec:sum} gives conclusions.


\section{Model}
\label{sec:model}

The chemical evolution model adopted in this work follows the framework of \citet{Kang2023}. In this model, the disk of M\,83 is assumed to form progressively through the accretion of primordial gas ($X = 0.7571$, $Y_{\rm p} = 0.2429$, $Z = 0$) from its surrounding halo. Outflow of metal-enriched gas is also incorporated in the model. The disk is divided into a set of independently evolved concentric rings, meaning that no radial mass flows are allowed in the model. 
In reality, radial flows can indeed occur in galaxies \citep[e.g. ][]{Sellwood1993, Ferguson2001, Rovskar2008, Spitoni2011, Kubryk2015, Calura2023, Kang2025}. 
Another simplification of the model is the omission of a bulge component---as the model is specifically developed to describe the star-forming disk rather than the bar/bulge---and the lack of differentiation between the thin and thick disks \citep[e.g. ][]{Spitoni2019}. 
Incorporating all these phenomena analytically is far from trivial and would entail numerous additional free parameters, the values of which might be difficult to constrain.
Despite these simplifying assumptions, the model has been successfully applied to the study of disk formation and evolution in the Milky Way and other nearby disk galaxies \citep[e.g. ][]{BP2000, MM2011, Bouquin2018, Bresolin2019, Belfiore2019, Kang2015, Kang2023}.
Consequently, the main physical processes incorporated in the model are: the infall of metal-free gas, a star formation law, and outflows of metal-enriched gas. 


To explain the relative scarcity of observed low-metallicity stars (the G-dwarf problem) in galaxy disks, the model invokes a continuous infall of primordial gas \citep[e.g. ][]{vandenBergh1962, Haywood2019, Matteucci2021}. The infall rate at radius $r$ and time $t$, $f_{\rm in}(r,t)$ (in units of $\rm M_\odot\,pc^{-2}\,Gyr^{-1}$), is expressed as:
\begin{equation}
f_{\rm{in}}(r,t) = A(r) \cdot t \cdot e^{-t/\tau},
\label{eq:infall rate}
\end{equation}
where $\tau$ is the gas infall timescale, and it is a free parameter in the model. 
Although a simple exponential form for the gas infall rate is commonly adopted in many previous chemical evolution models \citep[e.g. ][]{Matteucci1989, Hou2000,Schonrich2009, Spitoni2020, Calura2023}, the form used in this work allows for a broader range of evolutionary scenarios \citep{Kang2016}.
The function $A(r)$ is determined iteratively such that the present-day radial profile of stellar mass surface density, $\Sigma_*(r,t_{\mathrm{g}})$, predicted by the model matches the observed radial profile of M\,83. Here $t_{\mathrm{g}}$ denotes the cosmic age and is fixed at $13.5\,\mathrm{Gyr}$ in our adopted cosmology.

For M\,83, the observed present-day $\Sigma_*$ deviates from a simple exponential profile; we therefore directly use the observed radial profile from \citet{Bresolin2016} for our model.  
The measurements are derived from mid‑infrared surface photometry (WISE band W1 at 3.4~$\mu$m and Spitzer/IRAC at 3.6~$\mu$m), with surface magnitudes converted into mass surface densities. These data cover the radial range $0 \le R/{\rm R}_{25} \le 1.15$. 
In the outer disk ($1.5 \le R/{\rm R}_{25} \le 3.0$), where $\Sigma_*$ falls below the detection limit, \citet{Bresolin2016} rely on the empirical finding of \citet{Bigiel2010}.
The resulting $\Sigma_*$ profile thus covers the full range $0 \le R/{\rm R}_{25} \le 3.0$ and serves as a key input to our model.


The star formation rate surface density $\Sigma_{\rm SFR}(r,t)$ (in units of $M_\odot\,\text{pc}^{-2}\,\text{Gyr}^{-1}$) quantifies the mass of stars formed per unit area per unit time at a given radius $r$ and time $t$. It describes how efficiently a disk galaxy converts its cold gas into stars, given the galaxy's cold gas mass and scale radius. Early studies by \cite{Leroy2008} and \cite{Bigiel2008} showed that on sub-kiloparsec scales, $\Sigma_{\rm SFR}(r,t)$ correlates much more strongly with H$_{2}$ mass surface density $\Sigma_{\rm H_2}(r, t)$ than with HI mass surface density $\Sigma_{\rm HI}(r, t)$ or the total gas surface density $\Sigma_{\rm tot} = \Sigma_{\rm H_2} + \Sigma_{\rm HI}$. In other words, star formation activity is most directly linked to the H$_{2}$ component of the ISM  \citep{Kennicutt2012}. Building on this, \citet{Krumholz2014} concluded in a review that the correlation between star formation and H$_2$ is fundamental. Consequently, the relation $\Sigma_{\rm SFR} \propto \Sigma_{\rm H_2}$ has been adopted in several disk galaxy modeling studies \citep[e.g. ][]{Lagos2011, Kubryk2015, Kang2016, Kang2025}. $\Sigma_{\rm SFR}(r,t)$ proportional to $\Sigma_{\rm H_2}(r,t)$ is also adopted in the present work:
\begin{equation}
\Sigma_{\rm SFR}(r,t) = \Sigma_{\rm{H_2}}(r,t) / t_{\rm dep},
\label{eq:h2sfr}
\end{equation}
where $t_{\rm dep} = 1.9\,\rm Gyr$ is the molecular gas depletion time 
derived by \citet{Leroy2008} for a sample of 12 spiral galaxies and
subsequently adopted by \citet{Chang2012} for UGC\,8802, which has a stellar
mass similar to that of M\,83. 
To compute $\Sigma_{\rm H_2}(r,t)$, we decompose the total gas mass surface density $\Sigma_{\rm tot}$ into atomic and molecular components following the same method as described in detail by \citet{Kang2023}.

The gas outflow rate is assumed to be proportional to the $\Sigma_{\rm SFR}(r,t)$ \citep{Recchi2008}:
\begin{equation}
f_{\rm out}(r,t) = b_{\rm out} \Sigma_{\rm SFR}(r,t),
\label{eq:outflow}
\end{equation}
where \(b_{\rm out}\) is the dimensionless outflow efficiency. For a massive galaxy such as M\,83 (with stellar mass \(M_{\ast}\sim 10^{10.55}\,\rm{M_{\odot}}\); \citealt{Bresolin2016}), outflows are expected to play only a minor role in chemical evolution \citep{Tremonti2004, Spitoni2021, Calura2023}. This is because in such massive disks, the influence of outflows on long-term evolution is generally weaker than that of gas infall, and most ejected material is later reaccreted \citep{Toyouchi2015, Christensen2018}. Following the mass-dependent calibration of \citet{Chang2010}, we therefore adopt a low outflow efficiency, $b_{\rm out}=0.0045$. 
This value is consistent with those obtained for
galaxies of comparable mass: \citet{Chang2012} found $b_{\rm out}=0.007$ for
UGC\,8802, and \citet{Kang2015} found $b_{\rm out}=0.004$ for M51a, so the value
adopted here lies within the range inferred for similar systems.

Regarding the chemical evolution of the galactic disk, we assume both the instantaneous recycling approximation (IRA) and instantaneous mixing of gas with stellar ejecta. Under the IRA, stars more massive than $1\,M_{\odot}$ die instantaneously, while those below this mass live indefinitely. This approximation is reasonable for elements produced predominantly by short-lived massive stars (e.g., oxygen), but less accurate for elements synthesized over longer timescales, such as nitrogen, carbon, and iron \citep{Vincenzo2016, Matteucci2021}. Because oxygen is the most abundant heavy element by mass and serves as the best proxy for the overall ISM metallicity, we use the oxygen abundance ($12+\log(\rm O/H)$) to represent the ISM metallicity of M~83 throughout this work. The solar value is taken from \citet{Asplund2009}: $12+\log(\rm O/H)_{\odot} = 8.69$. Consequently, the gas in any fixed radial ring is characterized by a unique composition at each epoch. We adopt the classical chemical evolution equations of \citet{Tinsley1980} with the return fraction $R = 0.289$ and the yield $y = 0.019$, following the \citet{Kroupa1993} IMF and the metallicity-averaged yields from \citet{Vincenzo2016} based on \citet{Romano2010}. We neglect any dependence of $R$ and $y$ on metallicity or time. Infalling gas is primordial ($Z_{\rm{in}}=0$); outflowing gas is assumed to carry the same metallicity as the ISM at the time the outflow is launched, $Z_{\rm{out}}(r,t)=Z(r,t)$ \citep{Kang2023}.




In summary, the gas infall timescale $\tau$ is the only free parameter in our model.

\section{Observations}
\label{sec:obv}
A robust chemical evolution model for M\,83 must simultaneously satisfy multiple observational constraints: the radial distributions of $\Sigma_{\rm H_2}$ and $\Sigma_{\rm HI}$, $\Sigma_{\rm SFR}$, and the radial distributions of metallicity for both the ISM and young stars. The gas and SFR surface densities directly constrain the gas reservoir and current star formation activity. In contrast, the radial distributions of metallicity record the galaxy's evolutionary history---shaped by primordial gas infall, star formation, stellar feedback, and metal-rich outflows---and serve as a sensitive diagnostic tool for chemical evolution models \citep[e.g. ][]{BP2000, Kudritzki2015, Kubryk2015, Bresolin2019, Belfiore2019, Calura2023, Kang2025}.

The details of the available observations used to constrain the model are described in the following.

\subsection{Atomic hydrogen gas mass surface density}
\label{sec:HI_mass}

The $\Sigma_{\rm HI}$ of M\,83 is obtained from the HI 21\,cm line emission map observed with the NRAO' Very Large Array (VLA) observations in THINGS survey \citep{Walter2008}, and the radial profiles of the $\Sigma_{\rm HI}$ of M\,83 are taken from \citet{Bresolin2016} and \citet{Casasola2017}. 
The HI mass ($M_{\rm HI}$) of M\,83 has been calculated to be $M_{\rm HI}\sim 1.70-9.55\times10^{9}\,\rm M_{\odot}$ \citep{Walter2008, Barnes2014, Koribalski2018, Leroy2021b}.

\subsection{Molecular hydrogen gas mass surface density}
\label{sec:H2_mass}

\begin{figure*}
  \centering
  \includegraphics[angle=0,scale=0.55]{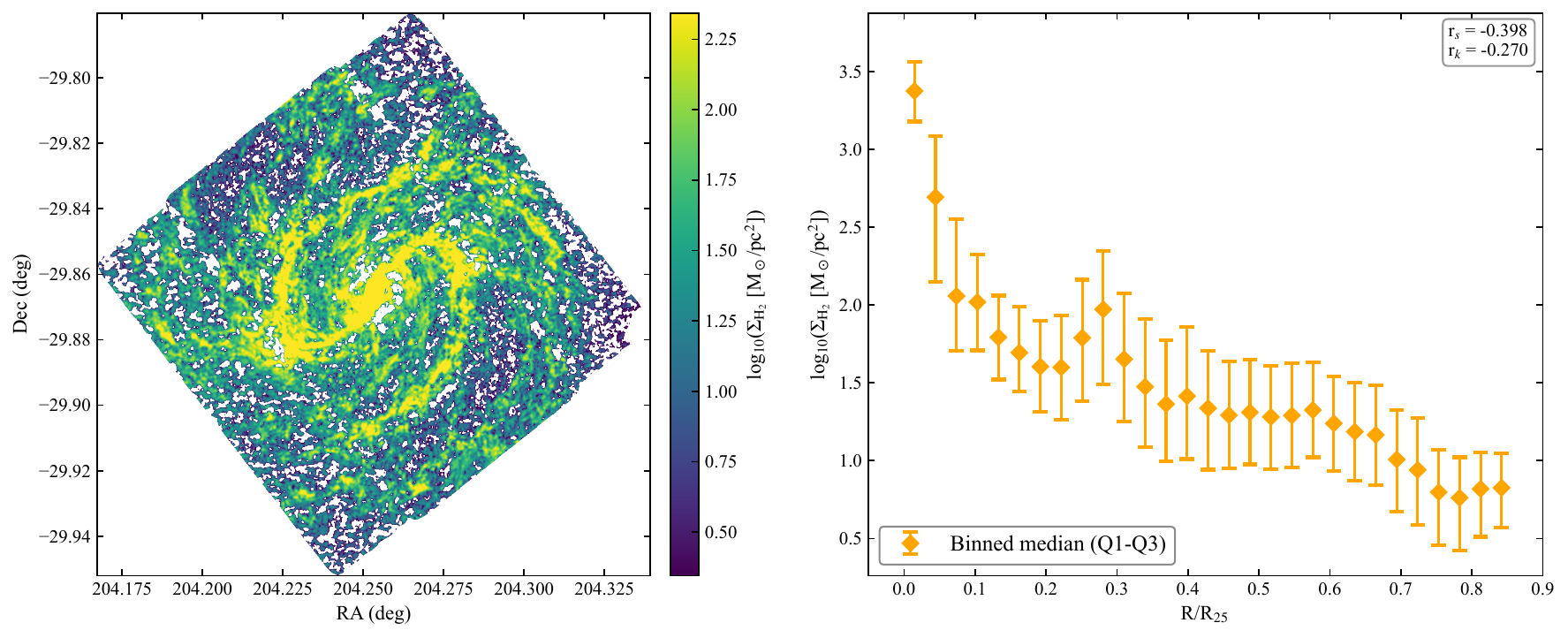}
    \caption{Two-dimensional and deprojected radial distributions of $\Sigma_{\rm H_{2}}$ on a logarithmic scale for M\,83. The median $\Sigma_{\rm H_{2}}$ and its interquartile range (Q1-Q3) within each radial bin are shown as orange diamonds with vertical error bars. 
}
\label{fig:H2}
\end{figure*}

The $\Sigma_{\rm H2}$ of M\,83 is derived from two sets of CO line emission maps.   
For the CO (1--0) line, we adopt the 115\,GHz map from \citet{Crosthwaite2002}, which was observed with the NRAO 12\,m telescope at Kitt Peak and subsequently processed by \citet{Bresolin2016}. We also adopt the ALMA CO (1--0) map processed by \citet{Koda2023}.
For the CO (2--1) line, we perform our own reduction and imaging of the ALMA data (see Figure.\,\ref{fig:H2} for the resulting map).

\begin{figure*}
  \centering
  \includegraphics[angle=0,scale=0.55]{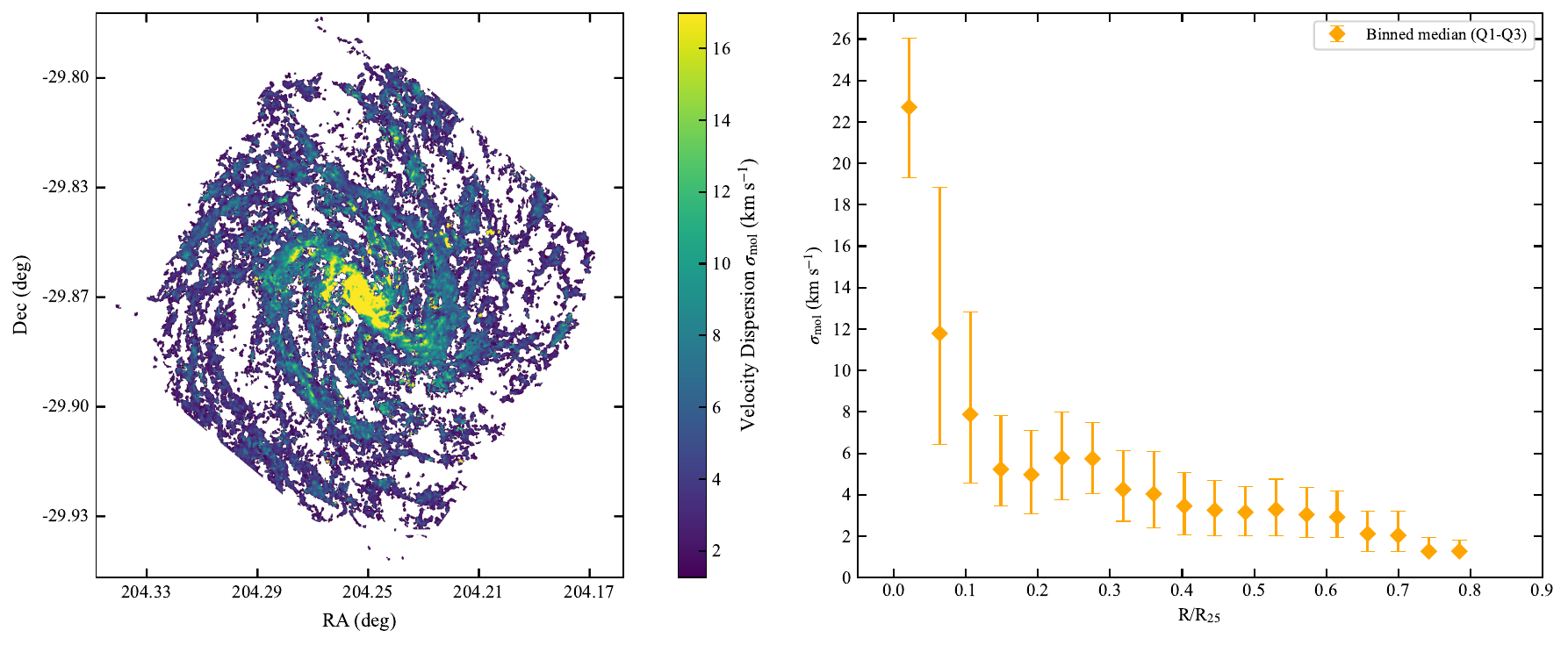}
    \caption{The two-dimensional and deprojected radial distributions 
    of velocity dispersion ($\sigma_v$) on a logarithmic scale for M\,83. The median $\Sigma_{\rm H_{2}}$ and its interquartile range (Q1-Q3) within each radial bin are shown as orange diamonds with vertical error bars. 
}
\label{fig:velocity_dispersion}
\end{figure*}

Using these CO (2--1) maps and following the PHANGS-ALMA methodology \citep{Leroy2021a,Leroy2021b}, we calculate $\Sigma_{\rm H2}$ for M\,83 from the CO (2--1) emission maps\footnote{\url{http://www.canfar.net/storage/list/phangs/RELEASES/PHANGS-ALMA/}}. 
The conversion proceeds in two steps. First, we convert the CO(2--1) intensity to CO(1--0) using a fixed line ratio of $0.65$ \citep{denBrok2021,Pessa2021}. 
Second, we convert the resulting CO(1--0) intensity to $\Sigma_{\rm H_{2}}$ using a CO-to-H$_2$ conversion factor $\alpha_{\rm CO}=4.35\ \rm M_\odot\,pc^{-2}\,(K\,km\,s^{-1})^{-1}$ \citep{Bolatto2013}. 
Adopting an inclination of $24^\circ$ ($\cos i = 0.9135$) and a helium correction factor of $1.36$, the overall conversion simplifies to $\Sigma_{\rm H_2}\ [\rm M_\odot\,pc^{-2}] = 3.51 \times I_{\rm CO(2-1)}\ [\rm K\,km\,s^{-1}]$.
The resulting radial profile of $\Sigma_{\rm H_{2}}$ presented in Figure\,\ref{fig:H2} exhibits a strong central concentration of molecular gas and falls rapidly at $R/\rm R_{25} \sim 0.8$. 
Such a concentration is consistent with previous observations of M\,83 from the ALMA CO (1--0) map \citep{Koda2023} and is frequently seen in barred spiral galaxies \citep[e.g. ][]{Matsuda1977, Sheth2005, Querejeta2021}.

The central region of M\,83 also displays an elevated velocity dispersion, as shown in Figure\,\ref{fig:velocity_dispersion}. The typical dispersion, corresponding to the peak of its probability density function, reaches approximately $\sigma_v \sim 23\, \rm {km\,s}^{-1}$ — markedly higher than that measured in other parts of the galaxy. A similar trend is observed in other barred spiral galaxies, where both the $\Sigma_{\rm H_2}$ and the velocity dispersion ($\sigma_v$) in the central region show an excess relative to the values found in their disks \citep{Sun2020}.
The H$_{2}$ gas mass of M\,83 is $M_{\rm H_{2}}\sim 1.9-7.08\times10^{9}\rm M_{\odot}$ \citep{Crosthwaite2002, Casasola2022, Koda2023, Lee2024}.

We note that $\alpha_{\rm CO}$ is not a universal constant but is known to vary with metallicity and local environmental conditions \citep{Sandstrom2013, Bolatto2013, Hunt2015, Teng2022, Teng2023}. Three
considerations justify adopting a single fiducial, Milky-Way-like value for M\,83. First, essentially all of the molecular gas is confined to the inner disk, with $\Sigma_{\rm H_2}$ dropping rapidly beyond $R/R_{25}\sim0.8$; in this region the ISM metallicity is close to or slightly above solar, which is
the regime in which the Galactic $\alpha_{\rm CO}$ is appropriate \citep{Sandstrom2013}. 
Second, although $\alpha_{\rm CO}$ is expected to be higher by a factor of a few in the metal-poor outer disk ($Z \sim 1/3\,Z_\odot$) \citep{Hunt2015, Bolatto2013}, the CO emission beyond $R/R_{25}\gtrsim1$ is weak and contributes negligibly to the total molecular mass. A metallicity-dependent $\alpha_{\rm CO}$ would therefore leave both the inner-disk $\Sigma_{\rm H_2}$ profile and the total $M_{\rm H_2}$ essentially unchanged. 
Third, resolved studies of nearby barred galaxies indicate that $\alpha_{\rm CO}$ can be lower by a factor of $\sim 2-3$ in bar-fed central molecular concentrations \citep{Sandstrom2013, Teng2022, Teng2023}; for M\,83, this would likely affect only the innermost radial bins and would lower the
central $\Sigma_{\rm H_2}$ peak accordingly. Even allowing for this systematic, its impact is confined to the central bins and is unlikely to change the qualitative model comparison; the predictions of viable theoretical models may still be able to reproduce the observations shown in panel 1 of Fig.\,\ref{fig:model_result}.


\subsection{Star formation rate surface density}
\label{sec:SFR}

The radial profiles of $\Sigma_{\rm SFR}$ of M\,83 are derived from two independent determinations: the \textit{GALEX} far-ultraviolet (FUV) emission with WISE $24\,\mu\rm m$ correction \citep{Barnes2014}, and the same FUV emission but corrected by WISE $22\,\mu\rm m$ data \citep{Casasola2017}.
The current total SFR in the disk of M\,83 is reported to be in the range $\sim1.6-6.96\,\rm M_{\odot }$\,yr$^{-1}$ using different tracers \citep{Jarrett2013, Barnes2014, Leroy2021b, Park2025}. 
Correspondingly, the observed sSFR, the ratio of SFR to stellar mass, can be calculated from ${\rm sSFR}\,=\,\frac{\rm SFR}{M_{*}}$, i.e., $\rm -10.346\,\leq\,log(sSFR/yr^{-1})\,\leq\,-9.869$.

\subsection{Metallicity of both the ISM and young stars}
\label{sec:H2}

The radial metallicity distributions along the disk of M\,83 have been estimated using several tracers, including BSGs \citep{Bresolin2016}, H{\sc ii} regions \citep{Bresolin2005, Bresolin2016, Grasha2022}, SSCs \citep{Davies2017}, YMCs
\citep{Hernandez2018, Hernandez2019, Hernandez2021}, and young stellar population metallicities from full-spectral fitting \citep{Sextl2025}. 
To constrain the model, we incorporate the observed radial metallicity profiles for M\,83, consisting of BSGs metallicity from \citet{Bresolin2016}, H{\sc ii} regions metallicity derived via the direct method and the N2O2 diagnostic \citep{Bresolin2005, Bresolin2016}, SSCs metallicity by \citet{Davies2017}, YMCs metallicities from \citet{Hernandez2018, Hernandez2019}, and the young stellar population metallicities obtained through full-spectral fitting of TYPHOON spectral data \citep{Sextl2025}. For completeness, we also include H{\sc ii} regions metallicity based on the strong line calibration of \citet{Dopita2016},  as measured by the TYPHOON survey \citep{Grasha2022}.


\section{Results and Discussion}
\label{sec:result_dis}

\subsection{Results}
\label{subsec:result}

\begin{figure}
  \centering
  \includegraphics[angle=0,scale=0.9]{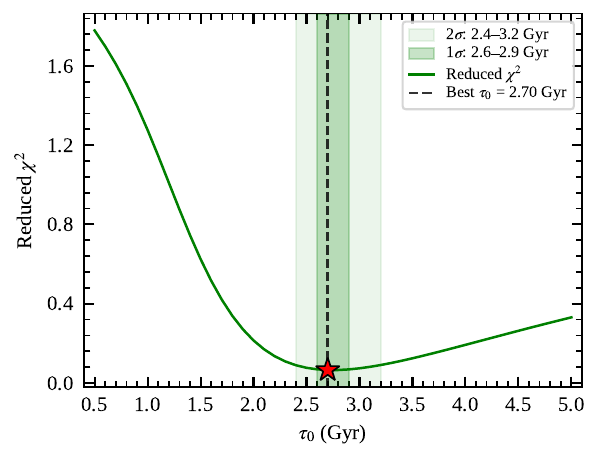}
    \caption{Reduced $\chi^2$ as a function of the infall timescale $\tau$. The green curve shows the combined reduced $\chi^2$ for the four observational datasets analyzed in this work. The dark and light green shaded regions indicate the $1\sigma$ and $2\sigma$ confidence intervals, respectively. The black dashed vertical line and the red star mark the best-fit value, located at the minimum of the curve.
    }
  \label{fig:reduced-chi2-tau}
\end{figure}

\begin{figure*}
  \centering
  \includegraphics[angle=0,scale=0.9]{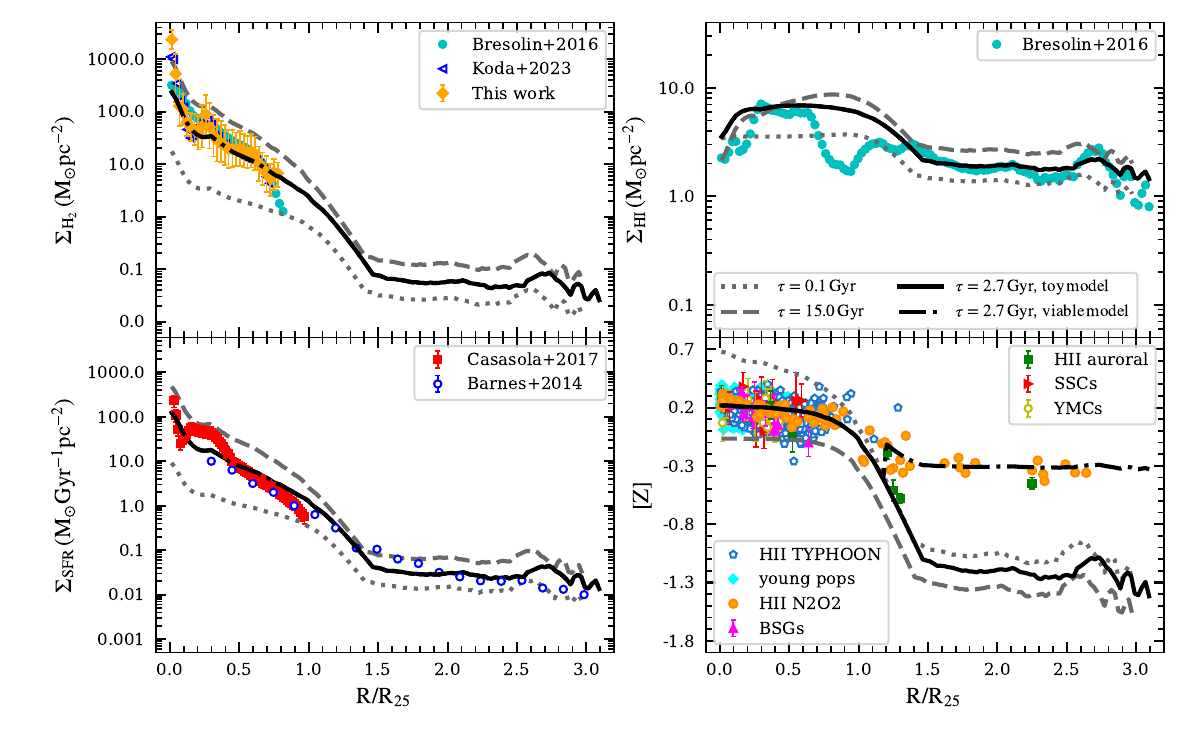}
    \caption{Comparison between model predictions and observed data of M\,83. 
The left panels show the radial profiles of H$_{2}$ mass (top) and SFR (bottom) surface densities. 
The right panels show the radial profiles of HI mass surface density (top) and metallicity [Z] for young stars and H{\sc ii} regions (bottom). 
The dotted and dashed lines correspond to the model results adopting $\tau = 0.1$\,Gyr and $\tau = 15$\,Gyr, respectively.
The  solid lines represent the results of the viable model, while the dash-dotted lines represent the results of the ``toy'' model.
Table\,\ref{tab:infall_params} presents the parameter set for our adopted solution for different line types: dotted, dashed, solid, and dash-dotted lines.
Observed data: 
H$_2$ data are shown as cyan filled circles \citep{Bresolin2016}, blue left open triangles \citep{Koda2023}, and orange filled diamonds (this work); 
SFR data are denoted by blue open circles \citep{Barnes2014} and red filled squares \citep{Casasola2017}; 
HI data are plotted as cyan filled circles \citep{Bresolin2016}. 
Metallicity data consist of: BSGs \citep[magenta solid upward triangles,][]{Bresolin2016}; 
H{\sc ii} regions from \citet{Bresolin2016} derived via the direct method (green filled squares) and the N2O2 diagnostic (orange filled circles); 
SSCs metallicities \citep[red rightward triangles,][]{Davies2017}; 
YMCs metallicities \citep[yellow open cycles,][]{Hernandez2018, Hernandez2019}; 
young stellar population metallicities from full-spectral fitting of TYPHOON data \citep[cyan filled diamonds,][]{Sextl2025}; 
and H{\sc ii} region metallicities based on the strong-line calibration of \citet{Dopita2016} as measured by the TYPHOON survey \citep[blue open pentagons,][]{Grasha2022}. To allow a direct comparison with the stellar metallicity, the oxygen abundances of the ISM are referenced to the solar value \(12 + \log(\rm{O/H}) = 8.69\) \citep{Asplund2009}. Note that metallicity data points with $R/R_{25} \lesssim 0.06$ are excluded from our analysis, as the central region is highly complex \citep{Harris2001, Callanan2021,Hernandez2025} and partially affected by recent localized infall of metal poor stellar or gaseous matter (see \citet{Sextl2025} and \citet{Sextl2026} for details). 
   }
\label{fig:model_result}
\end{figure*}

\begin{figure}
  \centering
  \includegraphics[angle=0,scale=0.78]{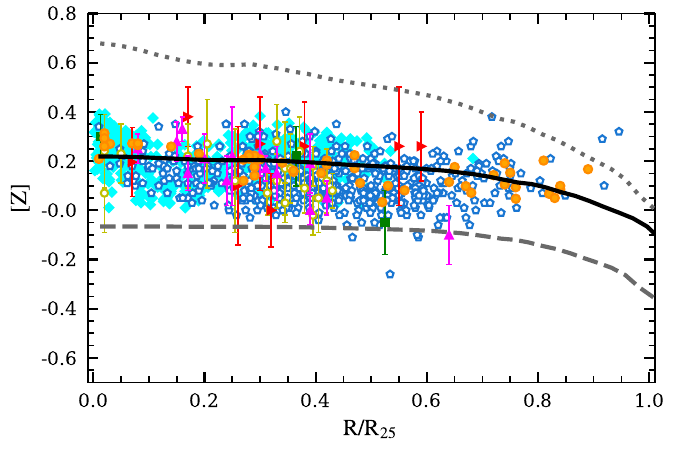}
    \caption{Observed metallicities and model fits across the M\,83 optical disk. The symbols, line-types and colors are the same as those in the lower-right panel of Figure \ref{fig:model_result}. }
  \label{fig:Zfit_large}
\end{figure}


To systematically investigate the influence of the free parameter $\tau$ (the gas infall timescale) on our model results, we examine two limiting cases: $\tau = 0.1\,\text{Gyr}$ and $\tau = 15\,\text{Gyr}$, shown as dotted and dashed lines, respectively, in Figure.\,\ref{fig:model_result}.
The left panels of Figure.\,\ref{fig:model_result} present the radial profiles of $\Sigma_{\rm H_{2}}$ (upper) and $\Sigma_{\rm SFR}$ (lower), while the right panels show the radial profiles of $\Sigma_{\rm HI}$ (upper) and $[Z]$ (lower).  
For $\Sigma_{\rm HI}$, $\Sigma_{\rm H_{2}}$, and $\Sigma_{\rm SFR}$, the area between the two model lines brackets the vast majority of the observed radial profiles across the entire disk of M\,83.  
Besides the few very central areas affected by very strong most recent metal poor infall (see Section \ref{sec:intro}), the same holds for the metallicity $[Z]$ in the inner region ($R/{\rm R_{25}} \lesssim 1.2$, see Figure \ref{fig:Zfit_large} for an enlarged view.). However, beyond $R/{\rm R_{25}} > 1.2$, 
both model predictions fall nearly one order of magnitude below the observed metallicities, indicating 
that under the present assumptions the model fails to reproduce the outer-disk chemical enrichment.

The region between the dotted and dashed lines indicates that an appropriate model (with a suitable $\tau$ between $0.1$ and $15\,\text{Gyr}$) exists that can explain all the observed features except for the metallicity distribution in the outer disk ($R/\rm{R_{25}} > 1.2$).
Guided by this parameter space exploration, we constrain the infall timescale $\tau$ by minimizing the reduced $\chi^2$ across four independent observational datasets: the $\rm{H}_2$ surface density profile, the $\rm{HI}$ surface density profile, the $\rm{SFR}$ surface density profile, and the metallicity profile. For the metallicity profile, we use only data with $R/R_{25} < 1.2$, whereas all available data are used for the other three profiles. The resulting curve is shown in Figure.\,\ref{fig:reduced-chi2-tau}. The minimum, marked by the red star, occurs at $\tau = 2.70\,\rm{Gyr}$, with $1\sigma$ and $2\sigma$ confidence intervals of $2.60$--$2.90\,\rm{Gyr}$ and $2.40$--$3.20\,\rm{Gyr}$, respectively. The reduced $\chi^2$ at the minimum is $0.07$, indicating an excellent fit to the data. This fitting procedure yields a viable model with $\tau = 2.70\,\rm{Gyr}$, which we adopt as our fiducial value in the subsequent analysis; its predictions are shown as the solid lines in Figure.\,\ref{fig:model_result}.
The model predictions reproduce the observed radial profiles of the H$_2$ and HI gas mass surface density and the SFR surface density with excellent agreement across the entire disk, including the extended HI disk and the low level of ongoing star formation.
For the metallicity, the model is in very good agreement with the observed data in the inner region of M\,83 ($R/{\rm R_{25}} \lesssim 1.2$), but fails to match the observed distribution in the outer disk ($R/{\rm R_{25}} \geq 1.2$). The outer disk metallicity discrepancy will be discussed further in a subsequent section.

\begin{figure*}
  \centering
  \includegraphics[angle=0,scale=0.88]{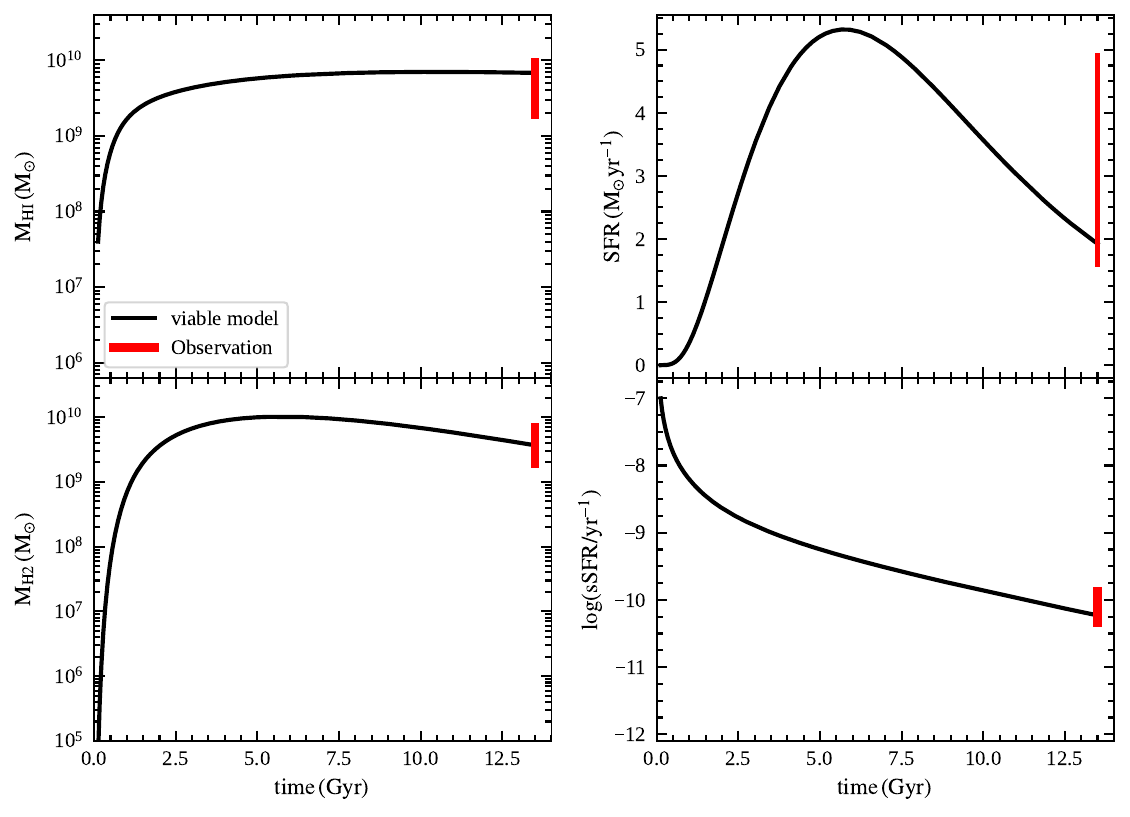}
    \caption{Time evolution of the global $\rm{M}_{\rm{HI}}$, $\rm{M}_{\rm{H2}}$, SFR, and sSFR of the M\,83 disk. The vertical red bars at 13.5\,Gyr indicate the observational constraints described in Section\,\ref{sec:obv}.
   }
  \label{fig:evolution}
\end{figure*}

The viable model predicts present-day values of $M_{{\rm H}_2} = 3.67 \times 10^{9} \, {\rm M}_{\odot}$, $M_{\rm{H I}} = 6.83 \times 10^{9} \, {\rm M}_{\odot}$, ${\rm SFR} = 1.93 \, {\rm M}_{\odot}\,{\rm yr}^{-1}$, and $\log({\rm sSFR} / {\rm yr}^{-1}) = -10.23$. 
These values are in good agreement with the observed ones described in Section\,\ref{sec:obv}, given the observational uncertainties. 
Figure\,\ref{fig:evolution} shows the temporal evolution of these four quantities from the beginning of the disk formation to the present day. The predictions of the viable model at the current epoch are in excellent agreement with the global observational data of the M\,83 disk, revealing a reasonable evolutionary path over the past $\sim13.5$\,Gyr. 
The core feature of this evolution is a cascading process of ``gas reservoir first, star formation later'': HI gas rapidly accumulates to $\sim10^9 {\rm M}_\odot$ in the early phase and then remains nearly constant; subsequently, HI is converted into H$_2$ gas, but supernova feedback significantly suppress the formation and survival of H$_2$, so that its mass peaks at $\sim5.0$--$5.5$ Gyr; the SFR follows the H$_2$ peak with a delay of $\sim1$ Gyr, reaching 
$\sim5 {\rm M}_\odot\,{\rm yr}^{-1}$, and then gradually declines as feedback dissipates the gas; 
the sSFR decreases monotonically with time from $\sim10^{-7}$ yr$^{-1}$ to $\sim10^{-10}$ yr$^{-1}$, reflecting the gradual aging of the galaxy.

\begin{figure}
  \centering
  \includegraphics[angle=0,scale=0.9]{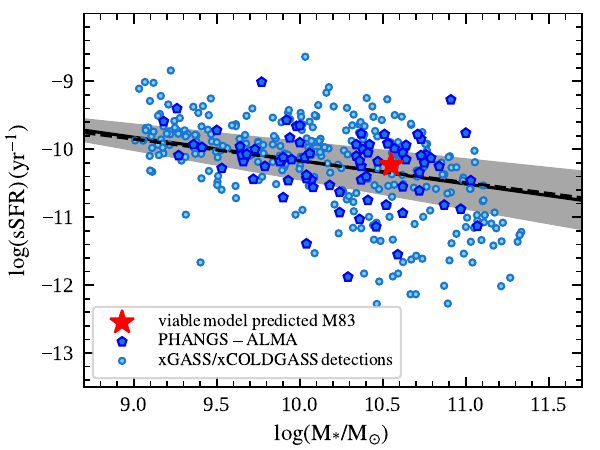}
    \caption{Specific star formation rate (sSFR) as a function of stellar mass. The red solid star shows the model-predicted sSFR for M\,83. The observed sSFRs from xCOLDGASS are shown as circles \citep{Catinella2018}, and those from PHANGS-ALMA are displayed as pentagons \citep{Leroy2021b}. The solid line represents the star-forming main sequence (SFMS), and the gray shaded region indicates the $1\sigma$ scatter \citep[eq.\,2]{Catinella2018}. The dashed line shows the SFMS using the specific formula from \citet{Leroy2019}.}
  \label{fig:Mstar_sSFR}
\end{figure}

\begin{figure}
  \centering
  \includegraphics[angle=0,scale=0.45]{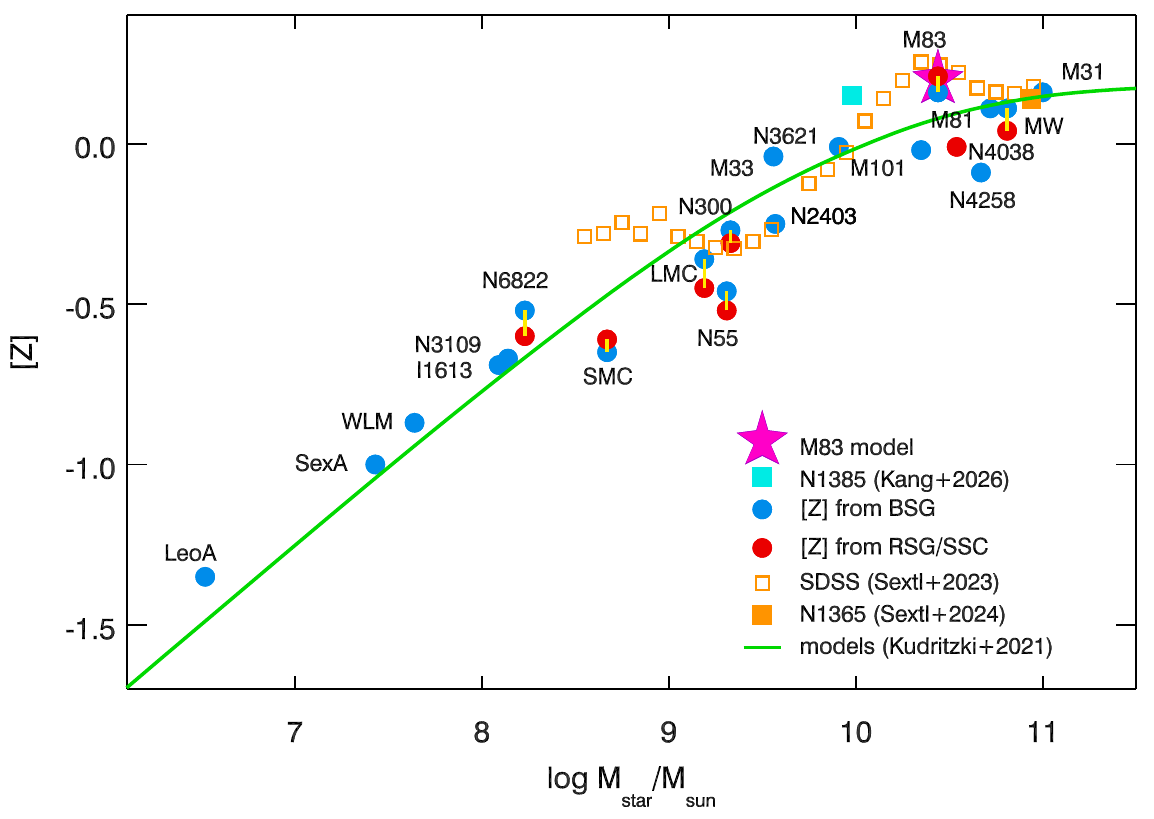}
    \caption{The mass--metallicity relationship of star-forming galaxies. Our M\,83 model prediction is shown as a pink solid star. Solid circles denote stellar metallicities derived from spectroscopy of individual blue supergiants (BSG, blue), red supergiants and super star clusters (RSG/SSC, red), \citep{Bresolin2025}. Open orange squares represent results from integrated galaxy spectra of 250,000 SDSS star-forming galaxies using a stellar population synthesis technique \citep{Sextl2023}, while the cyan and orange solid squares indicate the analysis results of spatially resolved young stellar population synthesis for NGC\,1385 \citep{Kang2026} and NGC\,1365 \citep{Sextl2024}, respectively. The prediction of a simple global galaxy evolution model \citep{Kudritzki2021a,Kudritzki2021b} is shown in green.
}
\label{fig:Mstar_Z}
\end{figure}

Beyond the evolutionary history described above, the viable model also successfully reproduces key present-day scaling relations. As shown in Figure.\,\ref{fig:Mstar_sSFR}, the predicted present-day sSFR, $\log(\text{sSFR} / \text{yr}^{-1}) = -10.23$, lies perfectly on the star-forming main sequence reported by \citet{Catinella2018} and \citet{Leroy2019}. Furthermore, Figure.\,\ref{fig:Mstar_Z} demonstrates that the viable model data point ($[Z]_{\rm M83} = +0.21$, the model-predicted present-day metallicity at a radius $r = 0.4R_{25}$) fits well with the mass--metallicity relation (MZR) of star-forming galaxies.

\begin{figure}
  \centering
  \includegraphics[angle=0,scale=0.89]{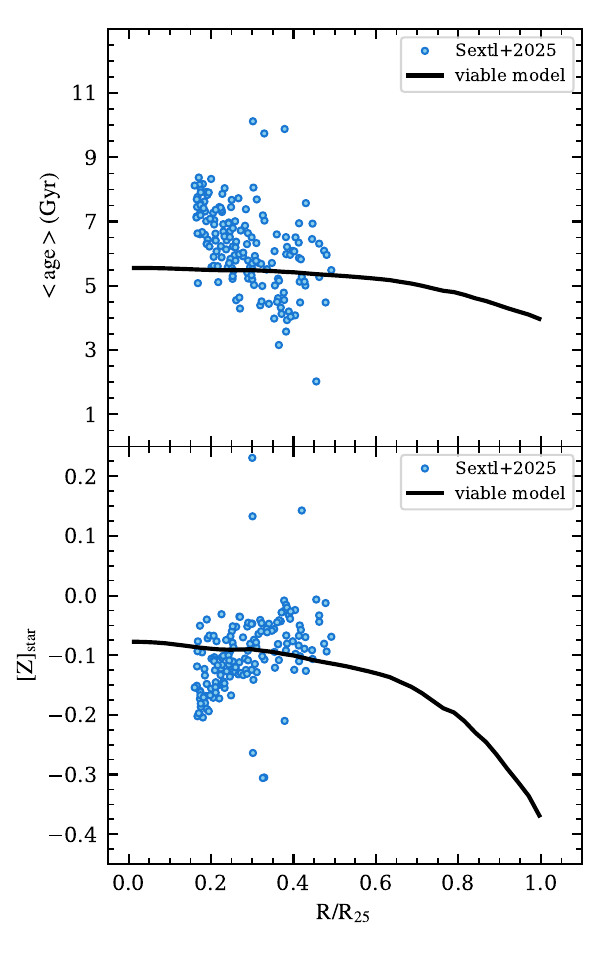}
    \caption{Radial model profiles of the mean stellar age (top) and mean stellar metallicity (bottom)  compared with the observed averages over all stars, young and old, as obtained by \citet{Sextl2025}.  
    }
  \label{fig:age_Z}
\end{figure}
 
We perform an additional test of the viable model by comparing its predictions with the observed mean metallicities and ages of the overall stellar population, which are derived from population synthesis techniques by \citet{Sextl2025}. For a given galactocentric radius $r$ and at the present epoch $t_{\rm g}$, the average mass-weighted stellar metallicity of the stellar population predicted by our model is computed as \citep{Kang2021}:
\begin{equation}
\langle\,Z_{\rm star}(r, t_{\rm g})\rangle\,=\,\frac{\int^{t_{\rm g}}_{0} Z_{\rm gas}(r,t')\,\Psi(r,t')\,{\rm d}t'}{\int^{t_{\rm g}}_{0} \Psi(r,t')\,{\rm d}t'},
\label{eq:Zstar}
\end{equation}
Similarly, the average mass-weighted logarithmic age of the stellar population is given by
\begin{equation}
\langle \log{\rm age}(r, t_{\rm g})\rangle = 
\frac{\int_{0}^{t_{\rm g}} \log(t_{\rm g}-t')\; \Psi(r,t')\; {\rm d}t'}
{\int_{0}^{t_{\rm g}} \Psi(r,t')\; {\rm d}t'}.
\label{eq:mean_log_age}
\end{equation}
In Figure\,\ref{fig:age_Z}, we present the results of this comparison, and the model agrees well with the observations in the radial range $0.15 \le R/{\rm R}_{25} \le 1.0$. 
The model predicts radially decreasing mean stellar age and metallicity: the inner disk is older and more metal-rich, while the outer disk is younger and more metal-poor. This inside-out formation scenario, as concluded from the model, implies that the inner region formed earlier and achieved more complete chemical enrichment.
In the region with $R/{\rm R}_{25} \lesssim 1.0$, the model-predicted metallicities of young stars and the ISM (bottom-right panel of Figure\,\ref{fig:model_result}) are approximately \(0.3\,{\rm dex}\) higher than the mean stellar metallicity (bottom panel of Figure\,\ref{fig:age_Z}). This offset is comparable to the young--old metallicity difference reported by \citet{Sextl2025}: in their Figure\,13, this difference averages around \(0.4\,{\rm dex}\) in the disk. Thus, both the present theoretical work and the observational work by \citet{Sextl2025} indicate that this offset reflects a significant degree of chemical evolution. 

\begin{figure}
  \centering
  \includegraphics[angle=0,scale=0.86]{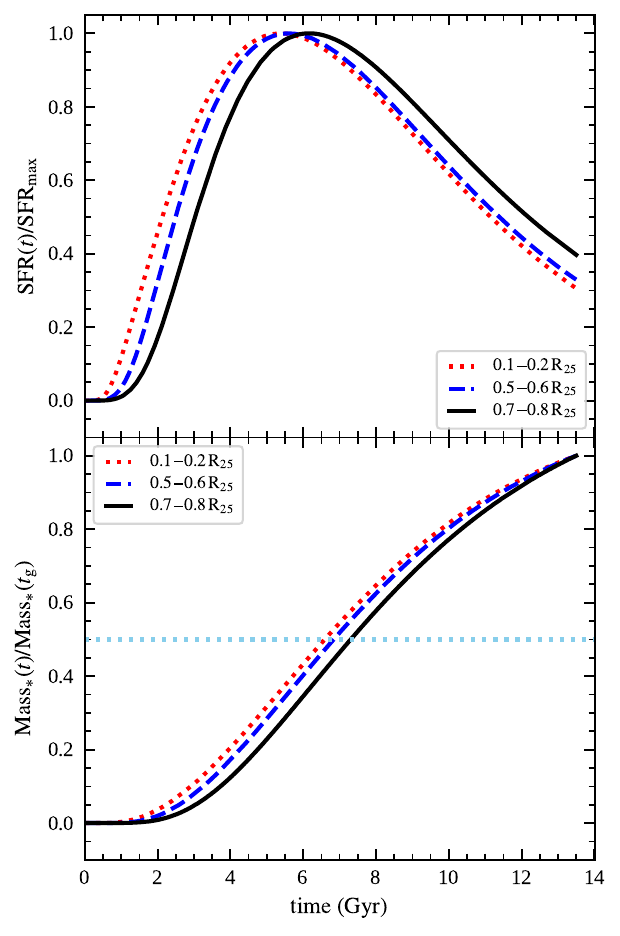}
    \caption{The viable model predicts the star formation histories (SFHs, top panel) and the relative growth of stellar mass (bottom panel) for three regions in the disk of M\,83, located at different galactocentric distances: $0.1$--$0.2\,{\rm R_{25}}$, $0.5$--$0.6\,{\rm R_{25}}$, and $0.7$--$0.8\,{\rm R_{25}}$. The SFH in each region is normalized to its peak value, while the stellar mass is normalized to its present-day value. The horizontal dotted line in the bottom panel marks the time at which each region reaches 50\% of its final stellar mass.
    }
  \label{fig:SFR_star_t}
\end{figure}

To examine the assembly of stellar populations across the M\,83 disk, Figure.\,\ref{fig:SFR_star_t} displays the viable model results for three radial bins, ranging from the inner disk ($0.1$--$0.2\,{\rm R_{25}}$) to the outer disk ($0.7$--$0.8\,{\rm R_{25}}$). 
The upper panel reveals a systematic delay in the peak of the SFH with increasing galactocentric distance: the innermost region peaks early, whereas the outermost region peaks later and exhibits a gradual decline. 
The lower panel shows that the inner region reaches 50\% of its present-day stellar mass much earlier than the outer region. Furthermore, the inner region grows rapidly at early epochs, while the outer region sustains a steadier growth over an extended period. This behavior indicates that mass assembly progresses from the inside out.
Notably, even though the model assumes a constant gas infall timescale, the predicted results in both Figure.\,\ref{fig:age_Z} and Figure.\,\ref{fig:SFR_star_t} successfully capture the inside-out formation of the M\,83 disk. Taken together, these features offer strong evidence for inside-out growth of the M\,83 disk, qualitatively consistent with the observational findings for M\,83 by \citet{Barnes2014}.


\begin{figure*}
  \centering
  \includegraphics[angle=0,scale=0.58]{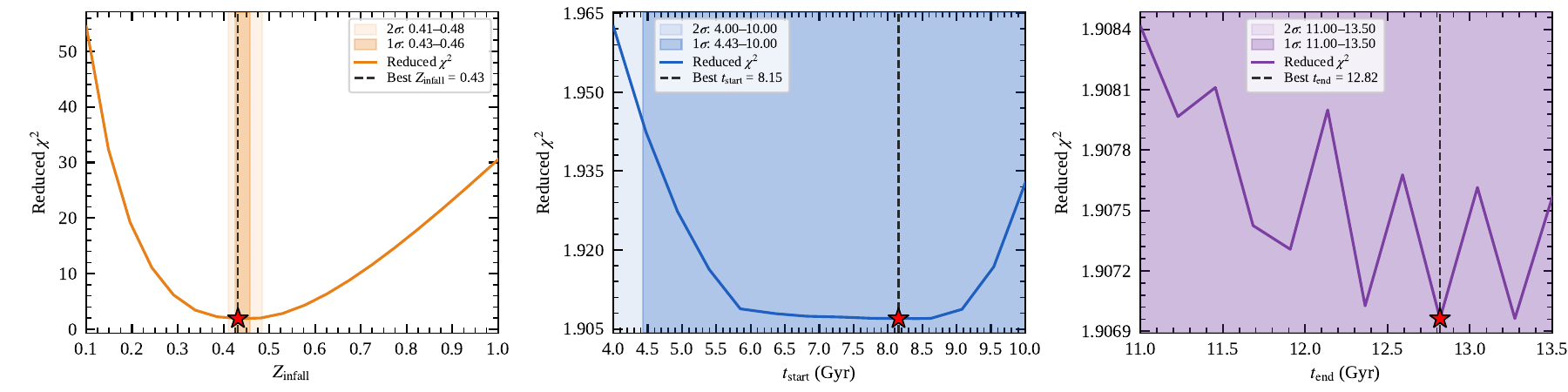}
    \caption{Reduced $\chi^2$ as a function of the ``toy'' model parameters: (left) the infall metallicity $Z_{\rm infall}$, (middle) the accretion start time $t_{\rm start}$, and (right) the accretion end time $t_{\rm end}$. The solid curves show the reduced $\chi^2$ for each parameter. The dark and light shaded regions indicate the $1\sigma$ and $2\sigma$ confidence intervals, respectively. The black dashed vertical line and the red star mark the best-fit value, located at the minimum of the curve.
    }
  \label{fig:reduced-chi2-out}
\end{figure*}

As noted earlier, the viable model fails to match the observed metallicity distribution in the outer disk ($R/\rm{R_{25}} > 1.2$). The key reason for this failure lies in the assumption that the gas in the outer disk starts from an initially metal-poor state and is enriched only through internal, inefficient star formation. This is essentially an in situ enrichment scenario, which, as shown by \citet{Kudritzki2014} for the extended disk of NGC\,3621, cannot explain the observed outer high metallicity over the disk's lifetime given the present low level of star formation. As argued by \citet{Bresolin2012} and \citet{Kudritzki2014}, the most viable mechanism to achieve such high metallicities is the ejection of enriched gas from central regions followed by its accretion onto the outer disk (i.e., outflow-enriched intergalactic gas accretion). The absence of these processes in our current model thus may cause the discrepancy in the outer disk.
Moreover, based on compelling evidence from cosmological simulations and observations, \citet{Bresolin2012} conclude that the outer disks of spiral galaxies are around 5\,Gyr in age.

To test this scenario, we present a ``toy'' model to simulate the accretion of pre-enriched cold gas onto the outer disk of M\,83 ($R/R_{25}\ge1.2$). All other model ingredients are identical to those of the viable model. The toy model introduces three additional parameters: the metallicity of the infalling gas ($Z_{\rm{infall_{out}}}$) and the start and end times of the pre-enriched accretion phase ($t_{\rm{start}}$ and $t_{\rm{end}}$). Rather than adopting these parameters by hand, we determine them through a systematic $\chi^2$ minimization against the observed outer-disk metallicity profile ($R/R_{25}\ge1.2$), following the same quantitative approach used to constrain $\tau$. We scan a three-dimensional grid over $Z_{\rm{infall_{out}}}/Z_\odot \in [0.1\text{--}1.0]$, $t_{\rm{start}}\in [4.0\text{--}10.0]\,\rm{Gyr}$ and $t_{\rm{end}}\in [11.0\text{--}13.5]\,\rm{Gyr}$; the resulting reduced $\chi^2$ curves are shown in Figure~\ref{fig:reduced-chi2-out}.

The $\chi^2$ surface exhibits a well-defined minimum at $Z_{\rm{infall_{out}}}/Z_\odot \simeq 0.43$ (with 1$\sigma$ and 2$\sigma$ ranges of 0.43--0.46 and 0.41--0.48, respectively). For the accretion onset, the best-fitting value is $t_{\rm{start}}/\rm{Gyr}\simeq 8.15$. In contrast, $\chi^2$ is essentially insensitive to $t_{\rm{end}}$, as the $\chi^2$ curve remains remarkably flat across the entire explored range. This is physically expected because the total accreted mass is fixed by the $\Sigma_*(r, t_{\rm g})$ normalization, and the declining infall rate at $t \gtrsim 8.15$\,Gyr limits the metals delivered in this late interval. Given that $Z_{\rm{infall_{out}}}$ and $t_{\rm{start}}$ are well constrained—as they directly set the metal budget and the onset of dilution—we therefore fix $t_{\rm{end}}=13.5$\,Gyr, i.e., the accretion of pre-enriched gas continues until the present day.

Therefore, our adopted ``toy'' model assumes that the outer disk accretes pre-enriched gas with a metallicity of $Z_{\rm{infall_{out}}}/Z_\odot=0.43$ from $t=8.15\,\rm{Gyr}$ until the present day ($t=13.5\,\rm{Gyr}$). We note that the constrained interval of $t_{\rm{start}}$ corresponds to lookback times of $\approx 5.35$\,Gyr, in excellent agreement with the $\sim5$\,Gyr age of the outer disks of spiral galaxies inferred by \citet{Bresolin2012}. Moreover, the best-fitting $Z_{\rm{infall_{out}}}$ lies slightly above the observed present-day outer-disk metallicity of $\sim1/3\,Z_\odot$ \citep{Bresolin2009}, as expected if the pre-enriched gas is diluted by the pre-existing metal-poor gas in the outer disk.
The results of the ``toy'' model shown by the dash-dotted line in Figure\,\ref{fig:model_result} fit the observed flat outer metallicity distribution very well. 
The parameters of our adopted solution are summarized in Table\,\ref{tab:infall_params}.
\begin{table}[H]
\caption{The parameters of our adopted model solution for different line types shown in Figure\,\ref{fig:model_result} and Figure\,\ref{fig:Zfit_large}.}
\centering
\label{tab:infall_params}
\begin{tabular}{lllll}
\hline
\hline
line-types  &  $R/{\rm R}_{25}$ & $t$ (Gyr) &  $\tau$ (Gyr) &  $Z_{\rm infall}$ \\
\hline
dotted line & 0--3.0            & 0--13.5   &   0.1         & 0 \\
dashed line & 0--3.0            & 0--13.5   &   15.0        & 0 \\
solid line  & 0--3.0            & 0--13.5   &   2.7         & 0 \\
\hline
dash-dotted line      & $< 1.2$           & 0--13.5         & 2.7    & 0 \\
dash-dotted line                      & $\ge 1.2$         & $0 \le t < 8.15$ & 2.7       & 0 \\
dash-dotted line                      & $\ge 1.2$         & $8.15 \le t \le 13.5$ & 2.7  & 0.43 \\
\hline
\end{tabular}
\end{table}

Our results demonstrate that the infall of pre-enriched gas in the outer disk successfully explains the extended, flat metal-rich HI disk observed in M\,83. We emphasize that, although the exact values of the free parameters in our model are not unique, the main conclusions regarding M\,83's SFH and the origin of its extended HI disk remain robust.


\subsection{Discussion}
\label{subsec:discussion}

Regarding the phenomenon of flat metallicity gradients in the outer disks of spiral galaxies, several potential causes have been proposed in the literature. These include flat gas surface densities in the outskirts \citep{Bresolin2009}, radial dependence of star-formation activity \citep{Phillipps1991}, gas redistribution caused by recent mergers \citep{Bresolin2009, Werk2011, Ogami2026}, a lack of evolution in extended galaxy disks \citep{Bresolin2009, Kudritzki2014}, the transport of metals to outer regions via the 'galactic fountain' effect \citep{Werk2011}, 
and radial gas flows \citep{Spitoni2011, Kubryk2015}. In addition, a recent study by \citet{Kewley2026} provides observational and simulation evidence that the flat metallicity gradient in NGC\,1365's outer disk was primarily shaped by a minor merger approximately 5--8\,Gyr ago (between redshifts $z=1$ and $z=0.5$). During this event, the galaxy accreted metal-rich gas from a dwarf galaxy, triggering the expansion of the galactic disk and directly shaping the chemical distribution of its outer regions.

In contrast to NGC\,1365, the case of M\,83 presents a different picture. \citet{Bell2026} used TNG-50 simulations to constrain M\,83's merger history and explore the origins of its nuclear starburst and extended HI disk. Their results show that while recent galaxy accretions can produce prominent stellar streams (including those known from earlier work \citep{Malin1997} and newly identified streams reported by \citet{Bell2026}), such events fail to trigger nuclear starburst activity and do not deliver enough gas to form the extended HI disk. Based on these findings, they conclude that merger events are unlikely to be the primary driver of M\,83's remarkable properties; instead, non-merger mechanisms such as secular evolution or gas accretion from the intergalactic medium (IGM) provide a more plausible explanation.

Although \citet{Bell2026} ruled out mergers as a direct means of delivering gas to form M\,83's extended HI disk based on TNG-50 simulations, merger events might still play an indirect role in gas accumulation. A more self-consistent possibility is that mergers induce tidal perturbations in the circumgalactic medium (CGM), which has been pre-enriched by M\,83's historical feedback activities. These perturbations trigger thermal instability and gas compression, leading to the cooling and precipitation of metal-enriched gas onto the outer disk. Such accretion would reshape the outer disk without introducing a steep radial metallicity gradient, because the infalling gas is uniformly enriched.

We note that radial gas flows are another mechanism often invoked to flatten metallicity gradients \citep[e.g. ][]{Spitoni2011, Kubryk2015}. However, such flows act as a continuous redistribution process and are expected to modify the gradient smoothly, whereas the observed M\,83 metallicity distribution exhibits a sharp break at $R/R_{25} \approx 1.0$ followed by a flat, metal-rich  outer disk. Indeed, the analytical accretion-disk model of \citet{Wang2022}, in which gas is transported radially through the disk, fails precisely because its solution cannot capture this discontinuity. Moreover, our viable model reproduces the inner-disk metallicity and the radial profiles of gas and SFR across the entire disk without radial flows, which argues against a dominant net radial transport in M\,83. Radial flows capable of enriching the outer disk would also require extracting substantial metal-rich gas from the inner disk, in tension with the well-reproduced inner profiles. In contrast, the pre-enriched infall scenario naturally explains both the extended HI disk and the flat outer metallicity.


\section{Conclusions}
\label{sec:sum}

In this work, we construct a simple and computationally efficient galactic chemical evolution model tailored to the nearby spiral galaxy M\,83, with the aim of investigating its complex assembly history and unusual radial metallicity distribution. The model assumes that the disk forms progressively via accretion of primordial gas, adopts a star formation law proportional to the molecular gas surface density, and includes metal-enriched outflows.  
The gas infall timescale, $\tau$, is the only free parameter. Guided by a systematic exploration of the parameter space, we perform a quantitative $\chi^{2}$ minimization against the observational constraints and identify a viable model with a best-fitting $\tau = 2.7$\,Gyr that successfully reproduces a comprehensive set of observational constraints. 
Specifically, the model accurately matches the observed radial profiles of HI mass surface density, H$_2$ mass surface density, and SFR surface density across the entire disk, while it reproduces the metallicity profiles of the ISM and young stars only within the inner disk ($R/R_{25} \lesssim 1.2$).

The model predictions for present-day global quantities are in excellent agreement with observations. These include $M_{\rm{H}_2} = 3.67\times 10^9\,M_\odot$, $M_{\rm{HI}} = 6.83\times 10^9\,M_\odot$, SFR $= 1.93\,M_\odot\,\rm{yr}^{-1}$, and $\log(\rm{sSFR}/\rm{yr}^{-1}) = -10.23$; in addition, the modeled galaxy lies perfectly on the star-forming main sequence and the mass--metallicity relation of star-forming galaxies. 
The reconstructed evolutionary history reveals a cascading process of ``gas reservoir first, star formation later'' over the past $\sim 13.5$\,Gyr, in which the SFR peaks with a delay of about $1$\,Gyr after the H$_2$ mass reaches its maximum. Furthermore, the model naturally captures an inside-out mass assembly history: inner regions form earlier and achieve more complete chemical enrichment, while predicting a $\sim 0.3$\,dex metallicity offset between young and total stellar populations, consistent with observational findings by \citet{Sextl2025}.

However, the viable model fails to reproduce the observed high metallicity in the outer disk ($R/R_{25} \ge 1.2$) under the assumptions of in situ chemical evolution, the current level of star formation, and initially metal-poor gas infall.
To resolve this discrepancy, we test a ``toy'' model in which pre-enriched gas with non-zero metallicity is accreted onto the outer disk. A systematic $\chi^2$ analysis yields best-fit values of $Z_{\rm infall}/Z_\odot \simeq 0.43$ for the infall metallicity, $8.15\,\mathrm{Gyr}$ for the accretion onset time, and $13.5\,\mathrm{Gyr}$ for the accretion termination time.
This scenario successfully explains the extended, flat, metal-rich HI disk observed in M\,83, demonstrating that the infall of pre-enriched gas, rather than in situ star formation, is the dominant enrichment mechanism in the outer regions. Finally, 
while the precise values of the free parameters in our model are not uniquely determined, our key findings concerning the SFH of M\,83 and the origin of its extended, flat, metal-rich HI disk remain  robust.

\begin{acknowledgements}
We thank the anonymous referee for constructive comments and suggestions, which led to significant improvements in the clarity and interpretation of our results.
This work is supported by the National Natural Science Foundation of China (NSFC Nos. 12288102, 12125303), the National Key R\&D Program of China (grant Nos. 2021YFA1600401, 2021YFA1600403), the Strategic Priority Research Program of the Chinese Academy of Sciences (grant Nos. XDB1160200, XDB1160201, XDB1160300, XDB1160303, XDB1160000), the Yunnan Revitalization Talent Support Program–Science \& Technology Champion Project (No. 202305AB350003), the New Cornerstone Science Foundation through the XPLORER PRIZE, the CAS Project for Young Scientists in Basic Research (YSBR-148) and International Centre of Supernovae (ICESUN), Yunnan Key Laboratory of Supernova Research (No. 202505AV340004), the Yunnan Revitalization Talent Support Program "YunLing Scholar" project.
Xiaoyu Kang thanks the support from the basic research program of Yunnan Province (No. 202401AT070142) and the Yunnan Revitalization Talent Support Program. Fenghui Zhang thanks the support from the NSF of China (No. 12573083). Eva Sextl, Rolf Kudritzki and Xiaoyu Kang acknowledge support by the Munich Excellence Cluster Origins and the Munich Instititute for Astro-, Particle and Biophysics (MIAPbP) both funded by the Deutsche Forschungsgemeinschaft (DFG, German Research Foundation) under the German Excellence Strategy EXC-2094 390783311. Xiaoyu Kang would like to thank Hye-Jin Park and Andrew J. Battisti for sharing data that supported this research. \\
This paper makes use of the following ALMA data: ADS/JAO.ALMA\#$2013.1.01161$.S,  ADS/JAO.ALMA\#$2015.1.00121$.S and ADS/JAO.ALMA\#$2016.1.00386$.S. 
ALMA is a partnership of ESO (representing its member states), NSF (USA) and NINS (Japan), together with NRC (Canada), MOST and ASIAA (Taiwan), and KASI (Republic of Korea), in cooperation with the Republic of Chile. The Joint ALMA Observatory is operated by ESO, AUI/NRAO and NAOJ. The National Radio Astronomy Observatory is a facility of the National Science Foundation operated under cooperative agreement by Associated Universities, Inc. 

\end{acknowledgements}

%
%

\bibliographystyle{aa} 
\bibliography{NGC5236} 

@ARTICLE{Kudritzki2021a,
       author = {{Kudritzki}, Rolf-Peter and {Teklu}, Adelheid F. and {Schulze}, Felix and {Remus}, Rhea-Silvia and {Dolag}, Klaus and {Burkert}, Andreas and {Zahid}, H. Jabran},
        title = "{Erratum: ``Galaxy Lookback Evolution Models: A Comparison with Magneticum Cosmological Simulations and Observations'' (2021, ApJ, 910 , 87)}",
      journal = {\apj},
         year = 2021,
        month = dec,
       volume = {922},
       number = {2},
          eid = {274},
        pages = {274},
          doi = {10.3847/1538-4357/ac32cf},
       adsurl = {https://ui.adsabs.harvard.edu/abs/2021ApJ...922..274K}
}

@ARTICLE{Kudritzki2021b,
       author = {{Kudritzki}, Rolf-Peter and {Teklu}, Adelheid F. and {Schulze}, Felix and {Remus}, Rhea-Silvia and {Dolag}, Klaus and {Burkert}, Andreas and {Zahid}, H. Jabran},
        title = "{Galaxy Look-back Evolution Models: A Comparison with Magneticum Cosmological Simulations and Observations}",
      journal = {\apj},
         year = 2021,
        month = apr,
       volume = {910},
       number = {2},
          eid = {87},
        pages = {87},
          doi = {10.3847/1538-4357/abe40c},
archivePrefix = {arXiv},
       eprint = {2102.04135},
 primaryClass = {astro-ph.GA},
       adsurl = {https://ui.adsabs.harvard.edu/abs/2021ApJ...910...87K}
}

@ARTICLE{Diaz2006,
       author = {{D{\'\i}az}, Rub{\'e}n J. and {Dottori}, Horacio and {Aguero}, Maria P. and {Mediavilla}, Evencio and {Rodrigues}, Irapuan and {Mast}, Damian},
        title = "{Hidden Trigger for the Giant Starburst Arc in M83?}",
      journal = {\apj},
         year = 2006,
        month = dec,
       volume = {652},
       number = {2},
        pages = {1122-1128},
          doi = {10.1086/507886},
archivePrefix = {arXiv},
       eprint = {astro-ph/0611771},
 primaryClass = {astro-ph},
       adsurl = {https://ui.adsabs.harvard.edu/abs/2006ApJ...652.1122D}
}

@ARTICLE{Chang2010,
       author = {{Chang}, R.~X. and {Hou}, J.~L. and {Shen}, S.~Y. and {Shu}, C.~G.},
        title = "{The Mass-dependent Star Formation Histories of Disk Galaxies: Infall Model Versus Observations}",
      journal = {\apj},
         year = 2010,
        month = oct,
       volume = {722},
       number = {1},
        pages = {380-387},
          doi = {10.1088/0004-637X/722/1/380},
archivePrefix = {arXiv},
       eprint = {1008.3045},
 primaryClass = {astro-ph.CO},
       adsurl = {https://ui.adsabs.harvard.edu/abs/2010ApJ...722..380C}
}

@ARTICLE{Davies2017,
       author = {{Davies}, Ben and {Kudritzki}, Rolf-Peter and {Lardo}, Carmela and {Bergemann}, Maria and {Beasor}, Emma and {Plez}, Bertrand and {Evans}, Chris and {Bastian}, Nate and {Patrick}, Lee R.},
        title = "{Red Supergiants as Cosmic Abundance Probes: Massive Star Clusters in M83 and the Mass-Metallicity Relation of Nearby Galaxies}",
      journal = {\apj},
         year = 2017,
        month = oct,
       volume = {847},
       number = {2},
          eid = {112},
        pages = {112},
          doi = {10.3847/1538-4357/aa89ed},
archivePrefix = {arXiv},
       eprint = {1708.08948},
 primaryClass = {astro-ph.GA},
       adsurl = {https://ui.adsabs.harvard.edu/abs/2017ApJ...847..112D}
}

@ARTICLE{Bresolin2005,
       author = {{Bresolin}, F. and {Schaerer}, D. and {Gonz{\'a}lez Delgado}, R.~M. and {Stasi{\'n}ska}, G.},
        title = "{A VLT study of metal-rich extragalactic H II regions. I. Observations and empirical abundances}",
      journal = {\aap},
         year = 2005,
        month = oct,
       volume = {441},
       number = {3},
        pages = {981-997},
          doi = {10.1051/0004-6361:20053369},
archivePrefix = {arXiv},
       eprint = {astro-ph/0506088},
 primaryClass = {astro-ph},
       adsurl = {https://ui.adsabs.harvard.edu/abs/2005A&A...441..981B}
}

@ARTICLE{Dopita2016,
       author = {{Dopita}, Michael A. and {Kewley}, Lisa J. and {Sutherland}, Ralph S. and {Nicholls}, David C.},
        title = "{Chemical abundances in high-redshift galaxies: a powerful new emission line diagnostic}",
      journal = {\apss},
         year = 2016,
        month = feb,
       volume = {361},
          eid = {61},
        pages = {61},
          doi = {10.1007/s10509-016-2657-8},
archivePrefix = {arXiv},
       eprint = {1601.01337},
 primaryClass = {astro-ph.GA},
       adsurl = {https://ui.adsabs.harvard.edu/abs/2016Ap&SS.361...61D}
}

@ARTICLE{Grasha2022,
       author = {{Grasha}, K. and {Chen}, Q.~H. and {Battisti}, A.~J. and {Acharyya}, A. and {Ridolfo}, S. and {Poehler}, E. and {Mably}, S. and {Verma}, A.~A. and {Hayward}, K.~L. and {Kharbanda}, A. and {Poetrodjojo}, H. and {Seibert}, M. and {Rich}, J.~A. and {Madore}, B.~F. and {Kewley}, L.~J.},
        title = "{Metallicity, Ionization Parameter, and Pressure Variations of H II Regions in the TYPHOON Spiral Galaxies: NGC 1566, NGC 2835, NGC 3521, NGC 5068, NGC 5236, and NGC 7793}",
      journal = {\apj},
         year = 2022,
        month = apr,
       volume = {929},
       number = {2},
          eid = {118},
        pages = {118},
          doi = {10.3847/1538-4357/ac5ab2},
archivePrefix = {arXiv},
       eprint = {2203.02522},
 primaryClass = {astro-ph.GA},
       adsurl = {https://ui.adsabs.harvard.edu/abs/2022ApJ...929..118G}
}

@ARTICLE{Bigiel2010,
       author = {{Bigiel}, F. and {Leroy}, A. and {Walter}, F. and {Blitz}, L. and {Brinks}, E. and {de Blok}, W.~J.~G. and {Madore}, B.},
        title = "{Extremely Inefficient Star Formation in the Outer Disks of Nearby Galaxies}",
      journal = {\aj},
         year = 2010,
        month = nov,
       volume = {140},
       number = {5},
        pages = {1194-1213},
          doi = {10.1088/0004-6256/140/5/1194},
archivePrefix = {arXiv},
       eprint = {1007.3498},
 primaryClass = {astro-ph.CO},
       adsurl = {https://ui.adsabs.harvard.edu/abs/2010AJ....140.1194B}
}

@ARTICLE{Kang2015,
   author = {{Kang}, X. and {Chang}, R. and {Zhang}, F. and {Cheng}, L. and
	{Wang}, L.},
    title = "{The evolution of interacting spiral galaxy NGC 5194}",
  journal = {\mnras},
archivePrefix = "arXiv",
   eprint = {1503.00398},
     year = 2015,
    month = may,
   volume = 449,
    pages = {414-424},
      doi = {10.1093/mnras/stv340},
   adsurl = {http://adsabs.harvard.edu/abs/2015MNRAS.449..414K}
}

@ARTICLE{Kang2023,
       author = {{Kang}, Xiaoyu and {Kudritzki}, Rolf-Peter and {Zhang}, Fenghui},
        title = "{The growth history of local M 33-mass bulgeless spiral galaxies}",
      journal = {\aap},
         year = 2023,
        month = nov,
       volume = {679},
          eid = {A83},
        pages = {A83},
          doi = {10.1051/0004-6361/202347677},
archivePrefix = {arXiv},
       eprint = {2309.07480},
 primaryClass = {astro-ph.GA},
       adsurl = {https://ui.adsabs.harvard.edu/abs/2023A&A...679A..83K}
}

@ARTICLE{Leroy2008,
   author = {{Leroy}, A.~K. and {Walter}, F. and {Brinks}, E. and {Bigiel}, F. and
	{de Blok}, W.~J.~G. and {Madore}, B. and {Thornley}, M.~D.},
    title = "{The Star Formation Efficiency in Nearby Galaxies: Measuring Where Gas Forms Stars Effectively}",
  journal = {\aj},
archivePrefix = "arXiv",
   eprint = {0810.2556},
     year = 2008,
    month = dec,
   volume = 136,
    pages = {2782-2845},
      doi = {10.1088/0004-6256/136/6/2782},
   adsurl = {http://adsabs.harvard.edu/abs/2008AJ....136.2782L}
}

@ARTICLE{Kroupa1993,
   author = {{Kroupa}, P. and {Tout}, C.~A. and {Gilmore}, G.},
    title = "{The distribution of low-mass stars in the Galactic disc}",
  journal = {\mnras},
     year = 1993,
    month = jun,
   volume = 262,
    pages = {545-587},
   adsurl = {http://adsabs.harvard.edu/abs/1993MNRAS.262..545K}
}

@ARTICLE{Asplund2009,
   author = {{Asplund}, M. and {Grevesse}, N. and {Sauval}, A.~J. and {Scott}, P.
	},
    title = "{The Chemical Composition of the Sun}",
  journal = {\araa},
archivePrefix = "arXiv",
   eprint = {0909.0948},
 primaryClass = "astro-ph.SR",
     year = 2009,
    month = sep,
   volume = 47,
    pages = {481-522},
      doi = {10.1146/annurev.astro.46.060407.145222},
   adsurl = {http://adsabs.harvard.edu/abs/2009ARA%26A..47..481A}
}

@ARTICLE{Matteucci1989,
   author = {{Matteucci}, F. and {Francois}, P.},
    title = "{Galactic chemical evolution - Abundance gradients of individual elements}",
  journal = {\mnras},
     year = 1989,
    month = aug,
   volume = 239,
    pages = {885-904},
   adsurl = {http://adsabs.harvard.edu/abs/1989MNRAS.239..885M}
}

@ARTICLE{Krumholz2014,
   author = {{Krumholz}, M.~R.},
    title = "{The big problems in star formation: The star formation rate, stellar clustering, and the initial mass function}",
  journal = {\physrep},
archivePrefix = "arXiv",
   eprint = {1402.0867},
     year = 2014,
    month = jun,
   volume = 539,
    pages = {49-134},
      doi = {10.1016/j.physrep.2014.02.001},
   adsurl = {http://adsabs.harvard.edu/abs/2014PhR...539...49K}
}

@ARTICLE{Querejeta2021,
       author = {{Querejeta}, M. and {Schinnerer}, E. and {Meidt}, S. and {Sun}, J. and {Leroy}, A.~K. and {Emsellem}, E. and {Klessen}, R.~S. and {Mu{\~n}oz-Mateos}, J.~C. and {Salo}, H. and {Laurikainen}, E. and {Be{\v{s}}li{\'c}}, I. and {Blanc}, G.~A. and {Chevance}, M. and {Dale}, D.~A. and {Eibensteiner}, C. and {Faesi}, C. and {Garc{\'\i}a-Rodr{\'\i}guez}, A. and {Glover}, S.~C.~O. and {Grasha}, K. and {Henshaw}, J. and {Herrera}, C. and {Hughes}, A. and {Kreckel}, K. and {Kruijssen}, J.~M.~D. and {Liu}, D. and {Murphy}, E.~J. and {Pan}, H. -A. and {Pety}, J. and {Razza}, A. and {Rosolowsky}, E. and {Saito}, T. and {Schruba}, A. and {Usero}, A. and {Watkins}, E.~J. and {Williams}, T.~G.},
        title = "{Stellar structures, molecular gas, and star formation across the PHANGS sample of nearby galaxies}",
      journal = {\aap},
         year = 2021,
        month = dec,
       volume = {656},
          eid = {A133},
        pages = {A133},
          doi = {10.1051/0004-6361/202140695},
archivePrefix = {arXiv},
       eprint = {2109.04491},
 primaryClass = {astro-ph.GA},
       adsurl = {https://ui.adsabs.harvard.edu/abs/2021A&A...656A.133Q}
}

@ARTICLE{Sextl2025,
       author = {{Sextl}, Eva and {Kudritzki}, Rolf-Peter and {Bresolin}, Fabio and {Grasha}, Kathryn and {Park}, Hye-Jin and {Chen}, Qian-Hui and {Battisti}, Andrew J. and {Seibert}, Mark and {Madore}, Barry F. and {Rich}, Jeffrey A.},
        title = "{The TYPHOON Stellar Population Synthesis Survey. II. Pushing Full Spectral Fitting to the Limit in the Nearby Grand Design Barred Spiral M83}",
      journal = {\apj},
         year = 2025,
        month = jul,
       volume = {987},
       number = {2},
          eid = {138},
        pages = {138},
          doi = {10.3847/1538-4357/addec2},
archivePrefix = {arXiv},
       eprint = {2505.21127},
 primaryClass = {astro-ph.GA},
       adsurl = {https://ui.adsabs.harvard.edu/abs/2025ApJ...987..138S}
}

@ARTICLE{Sextl2024,
       author = {{Sextl}, Eva and {Kudritzki}, Rolf-Peter and {Burkert}, Andreas and {Ho}, I. -Ting and {Zahid}, H. Jabran and {Seibert}, Mark and {Battisti}, Andrew J. and {Madore}, Barry F. and {Rich}, Jeffrey A.},
        title = "{The TYPHOON Stellar Population Synthesis Survey. I. The Young Stellar Population of the Great Barred Spiral NGC 1365}",
      journal = {\apj},
         year = 2024,
        month = jan,
       volume = {960},
       number = {1},
          eid = {83},
        pages = {83},
          doi = {10.3847/1538-4357/ad08b3},
archivePrefix = {arXiv},
       eprint = {2311.01140},
 primaryClass = {astro-ph.GA},
       adsurl = {https://ui.adsabs.harvard.edu/abs/2024ApJ...960...83S}
}

@ARTICLE{Pessa2021,
       author = {{Pessa}, I. and {Schinnerer}, E. and {Belfiore}, F. and {Emsellem}, E. and {Leroy}, A.~K. and {Schruba}, A. and {Kruijssen}, J.~M.~D. and {Pan}, H. -A. and {Blanc}, G.~A. and {Sanchez-Blazquez}, P. and {Bigiel}, F. and {Chevance}, M. and {Congiu}, E. and {Dale}, D. and {Faesi}, C.~M. and {Glover}, S.~C.~O. and {Grasha}, K. and {Groves}, B. and {Ho}, I. and {Jim{\'e}nez-Donaire}, M. and {Klessen}, R. and {Kreckel}, K. and {Koch}, E.~W. and {Liu}, D. and {Meidt}, S. and {Pety}, J. and {Querejeta}, M. and {Rosolowsky}, E. and {Saito}, T. and {Santoro}, F. and {Sun}, J. and {Usero}, A. and {Watkins}, E.~J. and {Williams}, T.~G.},
        title = "{Star formation scaling relations at {\ensuremath{\sim}}100 pc from PHANGS: Impact of completeness and spatial scale}",
      journal = {\aap},
         year = 2021,
        month = jun,
       volume = {650},
          eid = {A134},
        pages = {A134},
          doi = {10.1051/0004-6361/202140733},
archivePrefix = {arXiv},
       eprint = {2104.09536},
 primaryClass = {astro-ph.GA},
       adsurl = {https://ui.adsabs.harvard.edu/abs/2021A&A...650A.134P}
}

@ARTICLE{Bouquin2018,
       author = {{Bouquin}, Alexandre Y.~K. and {Gil de Paz}, Armando and {Mu{\~n}oz-Mateos}, Juan Carlos and {Boissier}, Samuel and {Sheth}, Kartik and {Zaritsky}, Dennis and {Peletier}, Reynier F. and {Knapen}, Johan H. and {Gallego}, Jes{\'u}s},
        title = "{The GALEX/S$^{4}$G Surface Brightness and Color Profiles Catalog. I. Surface Photometry and Color Gradients of Galaxies}",
      journal = {\apjs},
         year = 2018,
        month = feb,
       volume = {234},
       number = {2},
          eid = {18},
        pages = {18},
          doi = {10.3847/1538-4365/aaa384},
archivePrefix = {arXiv},
       eprint = {1710.00955},
 primaryClass = {astro-ph.GA},
       adsurl = {https://ui.adsabs.harvard.edu/abs/2018ApJS..234...18B}
}

@ARTICLE{Leroy2021a,
       author = {{Leroy}, Adam K. and {Hughes}, Annie and {Liu}, Daizhong and {Pety}, J{\'e}r{\^o}me and {Rosolowsky}, Erik and {Saito}, Toshiki and {Schinnerer}, Eva and {Schruba}, Andreas and {Usero}, Antonio and {Faesi}, Christopher M. and {Herrera}, Cinthya N. and {Chevance}, M{\'e}lanie and {Hygate}, Alexander P.~S. and {Kepley}, Amanda A. and {Koch}, Eric W. and {Querejeta}, Miguel and {Sliwa}, Kazimierz and {Will}, David and {Wilson}, Christine D. and {Anand}, Gagandeep S. and {Barnes}, Ashley and {Belfiore}, Francesco and {Be{\v{s}}li{\'c}}, Ivana and {Bigiel}, Frank and {Blanc}, Guillermo A. and {Bolatto}, Alberto D. and {Boquien}, M{\`e}d{\`e}ric and {Cao}, Yixian and {Chandar}, Rupali and {Chastenet}, J{\'e}r{\'e}my and {Chiang}, I-Da and {Congiu}, Enrico and {Dale}, Daniel A. and {Deger}, Sinan and {den Brok}, Jakob S. and {Eibensteiner}, Cosima and {Emsellem}, Eric and {Garc{\'\i}a-Rodr{\'\i}guez}, Axel and {Glover}, Simon C.~O. and {Grasha}, Kathryn and {Groves}, Brent and {Henshaw}, Jonathan D. and {Jim{\'e}nez Donaire}, Mar{\'\i}a J. and {Kim}, Jaeyeon and {Klessen}, Ralf S. and {Kreckel}, Kathryn and {Kruijssen}, J.~M. Diederik and {Larson}, Kirsten L. and {Lee}, Janice C. and {Mayker}, Ness and {McElroy}, Rebecca and {Meidt}, Sharon E. and {Mok}, Angus and {Pan}, Hsi-An and {Puschnig}, Johannes and {Razza}, Alessandro and {S{\'a}nchez-Bl'azquez}, Patricia and {Sandstrom}, Karin M. and {Santoro}, Francesco and {Sardone}, Amy and {Scheuermann}, Fabian and {Sun}, Jiayi and {Thilker}, David A. and {Turner}, Jordan A. and {Ubeda}, Leonardo and {Utomo}, Dyas and {Watkins}, Elizabeth J. and {Williams}, Thomas G.},
        title = "{PHANGS-ALMA Data Processing and Pipeline}",
      journal = {\apjs},
         year = 2021,
        month = jul,
       volume = {255},
       number = {1},
          eid = {19},
        pages = {19},
          doi = {10.3847/1538-4365/abec80},
archivePrefix = {arXiv},
       eprint = {2104.07665},
 primaryClass = {astro-ph.IM},
       adsurl = {https://ui.adsabs.harvard.edu/abs/2021ApJS..255...19L}
}

@ARTICLE{Leroy2021b,
       author = {{Leroy}, Adam K. and {Schinnerer}, Eva and {Hughes}, Annie and {Rosolowsky}, Erik and {Pety}, J{\'e}r{\^o}me and {Schruba}, Andreas and {Usero}, Antonio and {Blanc}, Guillermo A. and {Chevance}, M{\'e}lanie and {Emsellem}, Eric and {Faesi}, Christopher M. and {Herrera}, Cinthya N. and {Liu}, Daizhong and {Meidt}, Sharon E. and {Querejeta}, Miguel and {Saito}, Toshiki and {Sandstrom}, Karin M. and {Sun}, Jiayi and {Williams}, Thomas G. and {Anand}, Gagandeep S. and {Barnes}, Ashley T. and {Behrens}, Erica A. and {Belfiore}, Francesco and {Benincasa}, Samantha M. and {Be{\v{s}}li{\'c}}, Ivana and {Bigiel}, Frank and {Bolatto}, Alberto D. and {den Brok}, Jakob S. and {Cao}, Yixian and {Chandar}, Rupali and {Chastenet}, J{\'e}r{\'e}my and {Chiang}, I-Da and {Congiu}, Enrico and {Dale}, Daniel A. and {Deger}, Sinan and {Eibensteiner}, Cosima and {Egorov}, Oleg V. and {Garc{\'\i}a-Rodr{\'\i}guez}, Axel and {Glover}, Simon C.~O. and {Grasha}, Kathryn and {Henshaw}, Jonathan D. and {Ho}, I. -Ting and {Kepley}, Amanda A. and {Kim}, Jaeyeon and {Klessen}, Ralf S. and {Kreckel}, Kathryn and {Koch}, Eric W. and {Kruijssen}, J.~M. Diederik and {Larson}, Kirsten L. and {Lee}, Janice C. and {Lopez}, Laura A. and {Machado}, Josh and {Mayker}, Ness and {McElroy}, Rebecca and {Murphy}, Eric J. and {Ostriker}, Eve C. and {Pan}, Hsi-An and {Pessa}, Ismael and {Puschnig}, Johannes and {Razza}, Alessandro and {S{\'a}nchez-Bl{\'a}zquez}, Patricia and {Santoro}, Francesco and {Sardone}, Amy and {Scheuermann}, Fabian and {Sliwa}, Kazimierz and {Sormani}, Mattia C. and {Stuber}, Sophia K. and {Thilker}, David A. and {Turner}, Jordan A. and {Utomo}, Dyas and {Watkins}, Elizabeth J. and {Whitmore}, Bradley},
        title = "{PHANGS-ALMA: Arcsecond CO(2-1) Imaging of Nearby Star-forming Galaxies}",
      journal = {\apjs},
         year = 2021,
        month = dec,
       volume = {257},
       number = {2},
          eid = {43},
        pages = {43},
          doi = {10.3847/1538-4365/ac17f3},
archivePrefix = {arXiv},
       eprint = {2104.07739},
 primaryClass = {astro-ph.GA},
       adsurl = {https://ui.adsabs.harvard.edu/abs/2021ApJS..257...43L}
}

@ARTICLE{Vincenzo2016,
       author = {{Vincenzo}, F. and {Matteucci}, F. and {Belfiore}, F. and {Maiolino}, R.},
        title = "{Modern yields per stellar generation: the effect of the IMF}",
      journal = {\mnras},
         year = 2016,
        month = feb,
       volume = {455},
       number = {4},
        pages = {4183-4190},
          doi = {10.1093/mnras/stv2598},
archivePrefix = {arXiv},
       eprint = {1503.08300},
 primaryClass = {astro-ph.GA},
       adsurl = {https://ui.adsabs.harvard.edu/abs/2016MNRAS.455.4183V}
}

@ARTICLE{denBrok2021,
       author = {{den Brok}, J.~S. and {Chatzigiannakis}, D. and {Bigiel}, F. and {Puschnig}, J. and {Barnes}, A.~T. and {Leroy}, A.~K. and {Jim{\'e}nez-Donaire}, M.~J. and {Usero}, A. and {Schinnerer}, E. and {Rosolowsky}, E. and {Faesi}, C.~M. and {Grasha}, K. and {Hughes}, A. and {Kruijssen}, J.~M.~D. and {Liu}, D. and {Neumann}, L. and {Pety}, J. and {Querejeta}, M. and {Saito}, T. and {Schruba}, A. and {Stuber}, S.},
        title = "{New constraints on the $^{12}$CO(2-1)/(1-0) line ratio across nearby disc galaxies}",
      journal = {\mnras},
         year = 2021,
        month = jul,
       volume = {504},
       number = {3},
        pages = {3221-3245},
          doi = {10.1093/mnras/stab859},
archivePrefix = {arXiv},
       eprint = {2103.10442},
 primaryClass = {astro-ph.GA},
       adsurl = {https://ui.adsabs.harvard.edu/abs/2021MNRAS.504.3221D}
}

@ARTICLE{Bolatto2013,
       author = {{Bolatto}, Alberto D. and {Wolfire}, Mark and {Leroy}, Adam K.},
        title = "{The CO-to-H$_{2}$ Conversion Factor}",
      journal = {\araa},
         year = 2013,
        month = aug,
       volume = {51},
       number = {1},
        pages = {207-268},
          doi = {10.1146/annurev-astro-082812-140944},
archivePrefix = {arXiv},
       eprint = {1301.3498},
 primaryClass = {astro-ph.GA},
       adsurl = {https://ui.adsabs.harvard.edu/abs/2013ARA&A..51..207B}
}

@ARTICLE{Bresolin2025,
       author = {{Bresolin}, Fabio and {Kudritzki}, Rolf-Peter and {Urbaneja}, Miguel A. and {Sextl}, Eva and {Riess}, Adam G.},
        title = "{Blue Supergiants in the Pinwheel Galaxy M101: Comparison with H II Region Chemical Abundances, Spectroscopic Distance, and an Independent Determination of the Hubble Constant}",
      journal = {\apj},
         year = 2025,
        month = oct,
       volume = {991},
       number = {2},
          eid = {151},
        pages = {151},
          doi = {10.3847/1538-4357/adfc4c},
archivePrefix = {arXiv},
       eprint = {2508.11837},
 primaryClass = {astro-ph.GA},
       adsurl = {https://ui.adsabs.harvard.edu/abs/2025ApJ...991..151B}
}

@ARTICLE{Casasola2017,
       author = {{Casasola}, V. and {Cassar{\`a}}, L.~P. and {Bianchi}, S. and {Verstocken}, S. and {Xilouris}, E. and {Magrini}, L. and {Smith}, M.~W.~L. and {De Looze}, I. and {Galametz}, M. and {Madden}, S.~C. and {Baes}, M. and {Clark}, C. and {Davies}, J. and {De Vis}, P. and {Evans}, R. and {Fritz}, J. and {Galliano}, F. and {Jones}, A.~P. and {Mosenkov}, A.~V. and {Viaene}, S. and {Ysard}, N.},
        title = "{Radial distribution of dust, stars, gas, and star-formation rate in DustPedia face-on galaxies}",
      journal = {\aap},
         year = 2017,
        month = sep,
       volume = {605},
          eid = {A18},
        pages = {A18},
          doi = {10.1051/0004-6361/201731020},
archivePrefix = {arXiv},
       eprint = {1706.05351},
 primaryClass = {astro-ph.GA},
       adsurl = {https://ui.adsabs.harvard.edu/abs/2017A&A...605A..18C}
}

@ARTICLE{Park2025,
       author = {{Park}, Hye-Jin and {Battisti}, Andrew J. and {Marchal}, Antoine and {Cortese}, Luca and {Wisnioski}, Emily and {Seibert}, Mark and {Kim}, Shin-Jeong and {McClure-Griffiths}, Naomi and {de Blok}, W.~J.~G. and {Grasha}, Kathryn and {Madore}, Barry F. and {Rich}, Jeff A. and {Beaton}, Rachael L.},
        title = "{Exploring the potential for kinematically colder H I component as a tracer for star-forming gas in nearby galaxies}",
      journal = {\mnras},
         year = 2025,
        month = jul,
       volume = {540},
       number = {3},
        pages = {2623-2640},
          doi = {10.1093/mnras/staf902},
archivePrefix = {arXiv},
       eprint = {2506.01620},
 primaryClass = {astro-ph.GA},
       adsurl = {https://ui.adsabs.harvard.edu/abs/2025MNRAS.540.2623P}
}

@ARTICLE{Casasola2022,
       author = {{Casasola}, Viviana and {Bianchi}, Simone and {Magrini}, Laura and {Mosenkov}, Aleksandr V. and {Salvestrini}, Francesco and {Baes}, Maarten and {Calura}, Francesco and {Cassar{\`a}}, Letizia P. and {Clark}, Christopher J.~R. and {Corbelli}, Edvige and {Fritz}, Jacopo and {Galliano}, Fr{\'e}d{\'e}ric and {Liuzzo}, Elisabetta and {Madden}, Suzanne and {Nersesian}, Angelos and {Pozzi}, Francesca and {Roychowdhury}, Sambit and {Baronchelli}, Ivano and {Bonato}, Matteo and {Gruppioni}, Carlotta and {Pantoni}, Lara},
        title = "{The resolved scaling relations in DustPedia: Zooming in on the local Universe}",
      journal = {\aap},
         year = 2022,
        month = dec,
       volume = {668},
          eid = {A130},
        pages = {A130},
          doi = {10.1051/0004-6361/202245043},
archivePrefix = {arXiv},
       eprint = {2210.15993},
 primaryClass = {astro-ph.GA},
       adsurl = {https://ui.adsabs.harvard.edu/abs/2022A&A...668A.130C}
}

@ARTICLE{Bresolin2016,
       author = {{Bresolin}, Fabio and {Kudritzki}, Rolf-Peter and {Urbaneja}, Miguel A. and {Gieren}, Wolfgang and {Ho}, I.-Ting and {Pietrzy{\'n}ski}, Grzegorz},
        title = "{Young Stars and Ionized Nebulae in M83: Comparing Chemical Abundances at High Metallicity.}",
      journal = {\apj},
         year = 2016,
        month = oct,
       volume = {830},
       number = {2},
          eid = {64},
        pages = {64},
          doi = {10.3847/0004-637X/830/2/64},
archivePrefix = {arXiv},
       eprint = {1607.06840},
 primaryClass = {astro-ph.GA},
       adsurl = {https://ui.adsabs.harvard.edu/abs/2016ApJ...830...64B}
}

@ARTICLE{Crosthwaite2002,
       author = {{Crosthwaite}, Lucian P. and {Turner}, Jean L. and {Buchholz}, Leah and {Ho}, Paul T.~P. and {Martin}, Robert N.},
        title = "{CO in the Disk of the Barred Spiral Galaxy M83: CO (1-0), CO (2-1), and Neutral Gas}",
      journal = {\aj},
         year = 2002,
        month = apr,
       volume = {123},
       number = {4},
        pages = {1892-1912},
          doi = {10.1086/339479},
       adsurl = {https://ui.adsabs.harvard.edu/abs/2002AJ....123.1892C}
}

@ARTICLE{Koda2023,
       author = {{Koda}, Jin and {Hirota}, Akihiko and {Egusa}, Fumi and {Sakamoto}, Kazushi and {Sawada}, Tsuyoshi and {Heyer}, Mark and {Baba}, Junichi and {Boissier}, Samuel and {Calzetti}, Daniela and {Donovan Meyer}, Jennifer and {Elmegreen}, Bruce G. and {Gil de Paz}, Armando and {Harada}, Nanase and {Ho}, Luis C. and {Kobayashi}, Masato I.~N. and {Kuno}, Nario and {Lee}, Amanda M. and {Madore}, Barry F. and {Maeda}, Fumiya and {Mart{\'\i}n}, Sergio and {Muraoka}, Kazuyuki and {Nakanishi}, Kouichiro and {Onodera}, Sachiko and {Pineda}, Jorge L. and {Scoville}, Nick and {Watanabe}, Yoshimasa},
        title = "{Diverse Molecular Structures across the Whole Star-forming Disk of M83: High-fidelity Imaging at 40 pc Resolution}",
      journal = {\apj},
         year = 2023,
        month = jun,
       volume = {949},
       number = {2},
          eid = {108},
        pages = {108},
          doi = {10.3847/1538-4357/acc65e},
archivePrefix = {arXiv},
       eprint = {2303.12108},
 primaryClass = {astro-ph.GA},
       adsurl = {https://ui.adsabs.harvard.edu/abs/2023ApJ...949..108K}
}

@ARTICLE{Walter2008,
       author = {{Walter}, Fabian and {Brinks}, Elias and {de Blok}, W.~J.~G. and {Bigiel}, Frank and {Kennicutt}, Jr., Robert C. and {Thornley}, Michele D. and {Leroy}, Adam},
        title = "{THINGS: The H I Nearby Galaxy Survey}",
      journal = {\aj},
         year = 2008,
        month = dec,
       volume = {136},
       number = {6},
        pages = {2563-2647},
          doi = {10.1088/0004-6256/136/6/2563},
archivePrefix = {arXiv},
       eprint = {0810.2125},
 primaryClass = {astro-ph},
       adsurl = {https://ui.adsabs.harvard.edu/abs/2008AJ....136.2563W}
}

@ARTICLE{Koribalski2018,
       author = {{Koribalski}, B{\"a}rbel S. and {Wang}, Jing and {Kamphuis}, P. and {Westmeier}, T. and {Staveley-Smith}, L. and {Oh}, S.-H. and {L{\'o}pez-S{\'a}nchez}, {\'A}. R. and {Wong}, O.~I. and {Ott}, J. and {de Blok}, W.~J.~G. and {Shao}, L.},
        title = "{The Local Volume H I Survey (LVHIS)}",
      journal = {\mnras},
         year = 2018,
        month = aug,
       volume = {478},
       number = {2},
        pages = {1611-1648},
          doi = {10.1093/mnras/sty479},
archivePrefix = {arXiv},
       eprint = {1904.09648},
 primaryClass = {astro-ph.GA},
       adsurl = {https://ui.adsabs.harvard.edu/abs/2018MNRAS.478.1611K}
}

@ARTICLE{Wang2022,
       author = {{Wang}, Enci and {Lilly}, Simon J.},
        title = "{Gas-phase Metallicity Profiles of Star-forming Galaxies in the Modified Accretion Disk Framework}",
      journal = {\apj},
         year = 2022,
        month = apr,
       volume = {929},
       number = {1},
          eid = {95},
        pages = {95},
          doi = {10.3847/1538-4357/ac5e31},
archivePrefix = {arXiv},
       eprint = {2201.04151},
 primaryClass = {astro-ph.GA},
       adsurl = {https://ui.adsabs.harvard.edu/abs/2022ApJ...929...95W}
}

@ARTICLE{Matsuda1977,
       author = {{Matsuda}, T. and {Nelson}, A.~H.},
        title = "{A galactic vacuum cleaner?}",
      journal = {\nat},
         year = 1977,
        month = apr,
       volume = {266},
       number = {5603},
        pages = {607-608},
          doi = {10.1038/266607a0},
       adsurl = {https://ui.adsabs.harvard.edu/abs/1977Natur.266..607M}
}

@ARTICLE{Sheth2005,
       author = {{Sheth}, Kartik and {Vogel}, Stuart N. and {Regan}, Michael W. and {Thornley}, Michele D. and {Teuben}, Peter J.},
        title = "{Secular Evolution via Bar-driven Gas Inflow: Results from BIMA SONG}",
      journal = {\apj},
         year = 2005,
        month = oct,
       volume = {632},
       number = {1},
        pages = {217-226},
          doi = {10.1086/432409},
archivePrefix = {arXiv},
       eprint = {astro-ph/0505393},
 primaryClass = {astro-ph},
       adsurl = {https://ui.adsabs.harvard.edu/abs/2005ApJ...632..217S}
}

@ARTICLE{Romano2010,
       author = {{Romano}, D. and {Karakas}, A.~I. and {Tosi}, M. and {Matteucci}, F.},
        title = "{Quantifying the uncertainties of chemical evolution studies. II. Stellar yields}",
      journal = {\aap},
         year = 2010,
        month = nov,
       volume = {522},
          eid = {A32},
        pages = {A32},
          doi = {10.1051/0004-6361/201014483},
archivePrefix = {arXiv},
       eprint = {1006.5863},
 primaryClass = {astro-ph.GA},
       adsurl = {https://ui.adsabs.harvard.edu/abs/2010A&A...522A..32R}
}

@ARTICLE{Matteucci2021,
       author = {{Matteucci}, Francesca},
        title = "{Modelling the chemical evolution of the Milky Way}",
      journal = {\aapr},
         year = 2021,
        month = dec,
       volume = {29},
       number = {1},
          eid = {5},
        pages = {5},
          doi = {10.1007/s00159-021-00133-8},
archivePrefix = {arXiv},
       eprint = {2106.13145},
 primaryClass = {astro-ph.GA},
       adsurl = {https://ui.adsabs.harvard.edu/abs/2021A&ARv..29....5M}
}

@ARTICLE{Bresolin2009,
       author = {{Bresolin}, Fabio and {Ryan-Weber}, Emma and {Kennicutt}, Robert C. and {Goddard}, Quinton},
        title = "{The Flat Oxygen Abundance Gradient in the Extended Disk of M83}",
      journal = {\apj},
         year = 2009,
        month = apr,
       volume = {695},
       number = {1},
        pages = {580-595},
          doi = {10.1088/0004-637X/695/1/580},
archivePrefix = {arXiv},
       eprint = {0901.1127},
 primaryClass = {astro-ph.GA},
       adsurl = {https://ui.adsabs.harvard.edu/abs/2009ApJ...695..580B}
}

@ARTICLE{Bresolin2012,
       author = {{Bresolin}, Fabio and {Kennicutt}, Robert C. and {Ryan-Weber}, Emma},
        title = "{Gas Metallicities in the Extended Disks of NGC 1512 and NGC 3621. Chemical Signatures of Metal Mixing or Enriched Gas Accretion?}",
      journal = {\apj},
         year = 2012,
        month = may,
       volume = {750},
       number = {2},
          eid = {122},
        pages = {122},
          doi = {10.1088/0004-637X/750/2/122},
archivePrefix = {arXiv},
       eprint = {1203.0956},
 primaryClass = {astro-ph.CO},
       adsurl = {https://ui.adsabs.harvard.edu/abs/2012ApJ...750..122B}
}

@ARTICLE{Kang2025,
       author = {{Kang}, Xiaoyu and {Kudritzki}, Rolf-Peter and {Gong}, Xiaobo and {Zhang}, Fenghui},
        title = "{Evolution and star formation history of NGC 300 from a chemical evolution model with radial gas inflows}",
      journal = {\aap},
         year = 2025,
        month = sep,
       volume = {701},
          eid = {A2},
        pages = {A2},
          doi = {10.1051/0004-6361/202554108},
archivePrefix = {arXiv},
       eprint = {2507.10245},
 primaryClass = {astro-ph.GA},
       adsurl = {https://ui.adsabs.harvard.edu/abs/2025A&A...701A...2K}
}

@ARTICLE{Kang2016,
       author = {{Kang}, Xiaoyu and {Zhang}, Fenghui and {Chang}, Ruixiang and {Wang}, Lang and {Cheng}, Liantao},
        title = "{The star formation history of low-mass disk galaxies: A case study of NGC 300}",
      journal = {\aap},
         year = 2016,
        month = jan,
       volume = {585},
          eid = {A20},
        pages = {A20},
          doi = {10.1051/0004-6361/201527041},
archivePrefix = {arXiv},
       eprint = {1510.04768},
 primaryClass = {astro-ph.GA},
       adsurl = {https://ui.adsabs.harvard.edu/abs/2016A&A...585A..20K}
}

@ARTICLE{Sun2020,
       author = {{Sun}, Jiayi and {Leroy}, Adam K. and {Schinnerer}, Eva and {Hughes}, Annie and {Rosolowsky}, Erik and {Querejeta}, Miguel and {Schruba}, Andreas and {Liu}, Daizhong and {Saito}, Toshiki and {Herrera}, Cinthya N. and {Faesi}, Christopher and {Usero}, Antonio and {Pety}, J{\'e}r{\^o}me and {Kruijssen}, J.~M. Diederik and {Ostriker}, Eve C. and {Bigiel}, Frank and {Blanc}, Guillermo A. and {Bolatto}, Alberto D. and {Boquien}, M{\'e}d{\'e}ric and {Chevance}, M{\'e}lanie and {Dale}, Daniel A. and {Deger}, Sinan and {Emsellem}, Eric and {Glover}, Simon C.~O. and {Grasha}, Kathryn and {Groves}, Brent and {Henshaw}, Jonathan and {Jimenez-Donaire}, Maria J. and {Kim}, Jenny J. and {Klessen}, Ralf S. and {Kreckel}, Kathryn and {Lee}, Janice C. and {Meidt}, Sharon and {Sandstrom}, Karin and {Sardone}, Amy E. and {Utomo}, Dyas and {Williams}, Thomas G.},
        title = "{Molecular Gas Properties on Cloud Scales across the Local Star-forming Galaxy Population}",
      journal = {\apjl},
         year = 2020,
        month = sep,
       volume = {901},
       number = {1},
          eid = {L8},
        pages = {L8},
          doi = {10.3847/2041-8213/abb3be},
archivePrefix = {arXiv},
       eprint = {2009.01842},
 primaryClass = {astro-ph.GA},
       adsurl = {https://ui.adsabs.harvard.edu/abs/2020ApJ...901L...8S}
}

@ARTICLE{Bell2026,
       author = {{Bell}, Eric F. and {Harmsen}, Benjamin and {Cosby}, Matthew and {Price}, Paul A. and {Pearson}, Sarah and {Monachesi}, Antonela and {de Jong}, Roelof S. and {Souza}, Richard D' and {Gozman}, Katya and {Nibauer}, Jacob and {Busch}, Michael P. and {Bailin}, Jeremy and {Holwerda}, Benne W. and {Jang}, In Sung and {Smercina}, Adam},
        title = "{The Low-mass and Structured Stellar Halo of M83 Argues Against a Merger Origin for Its Starburst and Extended Neutral Hydrogen Disk}",
      journal = {\apj},
         year = 2026,
        month = feb,
       volume = {997},
       number = {2},
          eid = {153},
        pages = {153},
          doi = {10.3847/1538-4357/ae2c5e},
archivePrefix = {arXiv},
       eprint = {2512.11744},
 primaryClass = {astro-ph.GA},
       adsurl = {https://ui.adsabs.harvard.edu/abs/2026ApJ...997..153B}
}

@ARTICLE{Kang2021,
       author = {{Kang}, Xiaoyu and {Chang}, Ruixiang and {Kudritzki}, Rolf-Peter and {Gong}, Xiaobo and {Zhang}, Fenghui},
        title = "{Breaking the degeneracy between gas inflow and outflows with stellar metallicity: insights on M 101}",
      journal = {\mnras},
         year = 2021,
        month = apr,
       volume = {502},
       number = {2},
        pages = {1967-1973},
          doi = {10.1093/mnras/stab147},
archivePrefix = {arXiv},
       eprint = {2101.06833},
 primaryClass = {astro-ph.GA},
       adsurl = {https://ui.adsabs.harvard.edu/abs/2021MNRAS.502.1967K}
}

@ARTICLE{Jarrett2013,
       author = {{Jarrett}, T.~H. and {Masci}, F. and {Tsai}, C.~W. and {Petty}, S. and {Cluver}, M.~E. and {Assef}, Roberto J. and {Benford}, D. and {Blain}, A. and {Bridge}, C. and {Donoso}, E. and {Eisenhardt}, P. and {Koribalski}, B. and {Lake}, S. and {Neill}, James D. and {Seibert}, M. and {Sheth}, K. and {Stanford}, S. and {Wright}, E.},
        title = "{Extending the Nearby Galaxy Heritage with WISE: First Results from the WISE Enhanced Resolution Galaxy Atlas}",
      journal = {\aj},
         year = 2013,
        month = jan,
       volume = {145},
       number = {1},
          eid = {6},
        pages = {6},
          doi = {10.1088/0004-6256/145/1/6},
archivePrefix = {arXiv},
       eprint = {1210.3628},
 primaryClass = {astro-ph.CO},
       adsurl = {https://ui.adsabs.harvard.edu/abs/2013AJ....145....6J}
}

@ARTICLE{Barnes2014,
       author = {{Barnes}, Kate L. and {van Zee}, Liese and {Dale}, Daniel A. and {Staudaher}, Shawn and {Bullock}, James S. and {Calzetti}, Daniela and {Chandar}, Rupali and {Dalcanton}, Julianne J.},
        title = "{New Insights on the Formation and Assembly of M83 from Deep Near-infrared Imaging}",
      journal = {\apj},
         year = 2014,
        month = jul,
       volume = {789},
       number = {2},
          eid = {126},
        pages = {126},
          doi = {10.1088/0004-637X/789/2/126},
       adsurl = {https://ui.adsabs.harvard.edu/abs/2014ApJ...789..126B}
}

@ARTICLE{Leroy2019,
       author = {{Leroy}, Adam K. and {Sandstrom}, Karin M. and {Lang}, Dustin and {Lewis}, Alexia and {Salim}, Samir and {Behrens}, Erica A. and {Chastenet}, J{\'e}r{\'e}my and {Chiang}, I-Da and {Gallagher}, Molly J. and {Kessler}, Sarah and {Utomo}, Dyas},
        title = "{A z = 0 Multiwavelength Galaxy Synthesis. I. A WISE and GALEX Atlas of Local Galaxies}",
      journal = {\apjs},
         year = 2019,
        month = oct,
       volume = {244},
       number = {2},
          eid = {24},
        pages = {24},
          doi = {10.3847/1538-4365/ab3925},
archivePrefix = {arXiv},
       eprint = {1910.13470},
 primaryClass = {astro-ph.GA},
       adsurl = {https://ui.adsabs.harvard.edu/abs/2019ApJS..244...24L}
}

@ARTICLE{BP2000,
       author = {{Boissier}, S. and {Prantzos}, N.},
        title = "{Chemo-spectrophotometric evolution of spiral galaxies - II. Main properties of present-day disc galaxies}",
      journal = {\mnras},
         year = 2000,
        month = feb,
       volume = {312},
       number = {2},
        pages = {398-416},
          doi = {10.1046/j.1365-8711.2000.03133.x},
       adsurl = {https://ui.adsabs.harvard.edu/abs/2000MNRAS.312..398B}
}

@ARTICLE{Bresolin2019,
       author = {{Bresolin}, Fabio},
        title = "{Metallicity gradients in small and nearby spiral galaxies}",
      journal = {\mnras},
         year = 2019,
        month = sep,
       volume = {488},
       number = {3},
        pages = {3826-3843},
          doi = {10.1093/mnras/stz1947},
archivePrefix = {arXiv},
       eprint = {1907.05071},
 primaryClass = {astro-ph.GA},
       adsurl = {https://ui.adsabs.harvard.edu/abs/2019MNRAS.488.3826B}
}

@ARTICLE{Kudritzki2015,
       author = {{Kudritzki}, Rolf-Peter and {Ho}, I.-Ting and {Schruba}, Andreas and {Burkert}, Andreas and {Zahid}, H. Jabran and {Bresolin}, Fabio and {Dima}, Gabriel I.},
        title = "{The chemical evolution of local star-forming galaxies: radial profiles of ISM metallicity, gas mass, and stellar mass and constraints on galactic accretion and winds}",
      journal = {\mnras},
         year = 2015,
        month = jun,
       volume = {450},
       number = {1},
        pages = {342-359},
          doi = {10.1093/mnras/stv522},
archivePrefix = {arXiv},
       eprint = {1503.01503},
 primaryClass = {astro-ph.GA},
       adsurl = {https://ui.adsabs.harvard.edu/abs/2015MNRAS.450..342K}
}

@ARTICLE{Kubryk2015,
       author = {{Kubryk}, M. and {Prantzos}, N. and {Athanassoula}, E.},
        title = "{Evolution of the Milky Way with radial motions of stars and gas. I. The solar neighbourhood and the thin and thick disks}",
      journal = {\aap},
         year = 2015,
        month = aug,
       volume = {580},
          eid = {A126},
        pages = {A126},
          doi = {10.1051/0004-6361/201424171},
archivePrefix = {arXiv},
       eprint = {1412.0585},
 primaryClass = {astro-ph.GA},
       adsurl = {https://ui.adsabs.harvard.edu/abs/2015A&A...580A.126K}
}

@ARTICLE{Belfiore2019,
       author = {{Belfiore}, F. and {Vincenzo}, F. and {Maiolino}, R. and {Matteucci}, F.},
        title = "{From `bathtub' galaxy evolution models to metallicity gradients}",
      journal = {\mnras},
         year = 2019,
        month = jul,
       volume = {487},
       number = {1},
        pages = {456-474},
          doi = {10.1093/mnras/stz1165},
archivePrefix = {arXiv},
       eprint = {1903.05105},
 primaryClass = {astro-ph.GA},
       adsurl = {https://ui.adsabs.harvard.edu/abs/2019MNRAS.487..456B}
}

@ARTICLE{vandenBergh1962,
       author = {{van den Bergh}, S.},
        title = "{The frequency of stars with different metal abundances.}",
      journal = {\aj},
         year = 1962,
        month = oct,
       volume = {67},
        pages = {486-490},
          doi = {10.1086/108757},
       adsurl = {https://ui.adsabs.harvard.edu/abs/1962AJ.....67..486V}
}

@ARTICLE{Haywood2019,
       author = {{Haywood}, M. and {Snaith}, O. and {Lehnert}, M.~D. and {Di Matteo}, P. and {Khoperskov}, S.},
        title = "{Revisiting long-standing puzzles of the Milky Way: the Sun and its vicinity as typical outer disk chemical evolution}",
      journal = {\aap},
         year = 2019,
        month = may,
       volume = {625},
          eid = {A105},
        pages = {A105},
          doi = {10.1051/0004-6361/201834155},
archivePrefix = {arXiv},
       eprint = {1903.03188},
 primaryClass = {astro-ph.GA},
       adsurl = {https://ui.adsabs.harvard.edu/abs/2019A&A...625A.105H}
}

@ARTICLE{Calura2023,
       author = {{Calura}, Francesco and {Palla}, Marco and {Morselli}, Laura and {Spitoni}, Emanuele and {Casasola}, Viviana and {Verma}, Kuldeep and {Enia}, Andrea and {Meneghetti}, Massimo and {Bianchi}, Simone and {Pozzi}, Francesca and {Gruppioni}, Carlotta},
        title = "{A Bayesian chemical evolution model of the DustPedia galaxy M74}",
      journal = {\mnras},
         year = 2023,
        month = aug,
       volume = {523},
       number = {2},
        pages = {2351-2368},
          doi = {10.1093/mnras/stad1316},
archivePrefix = {arXiv},
       eprint = {2305.05680},
 primaryClass = {astro-ph.GA},
       adsurl = {https://ui.adsabs.harvard.edu/abs/2023MNRAS.523.2351C}
}

@ARTICLE{Spitoni2020,
       author = {{Spitoni}, E. and {Calura}, F. and {Mignoli}, M. and {Gilli}, R. and {Silva Aguirre}, V. and {Gallazzi}, A.},
        title = "{Connection between galactic downsizing and the most fundamental galactic scaling relations}",
      journal = {\aap},
         year = 2020,
        month = oct,
       volume = {642},
          eid = {A113},
        pages = {A113},
          doi = {10.1051/0004-6361/202037879},
archivePrefix = {arXiv},
       eprint = {2003.02347},
 primaryClass = {astro-ph.GA},
       adsurl = {https://ui.adsabs.harvard.edu/abs/2020A&A...642A.113S}
}

@ARTICLE{Spitoni2021,
       author = {{Spitoni}, E. and {Calura}, F. and {Silva Aguirre}, V. and {Gilli}, R.},
        title = "{The G-dwarf distribution in star-forming galaxies: A tug of war between infall and outflow}",
      journal = {\aap},
         year = 2021,
        month = apr,
       volume = {648},
          eid = {L5},
        pages = {L5},
          doi = {10.1051/0004-6361/202140836},
archivePrefix = {arXiv},
       eprint = {2103.10965},
 primaryClass = {astro-ph.GA},
       adsurl = {https://ui.adsabs.harvard.edu/abs/2021A&A...648L...5S}
}

@ARTICLE{Tremonti2004,
       author = {{Tremonti}, Christy A. and {Heckman}, Timothy M. and {Kauffmann}, Guinevere and {Brinchmann}, Jarle and {Charlot}, St{\'e}phane and {White}, Simon D.~M. and {Seibert}, Mark and {Peng}, Eric W. and {Schlegel}, David J. and {Uomoto}, Alan and {Fukugita}, Masataka and {Brinkmann}, Jon},
        title = "{The Origin of the Mass-Metallicity Relation: Insights from 53,000 Star-forming Galaxies in the Sloan Digital Sky Survey}",
      journal = {\apj},
         year = 2004,
        month = oct,
       volume = {613},
       number = {2},
        pages = {898-913},
          doi = {10.1086/423264},
archivePrefix = {arXiv},
       eprint = {astro-ph/0405537},
 primaryClass = {astro-ph},
       adsurl = {https://ui.adsabs.harvard.edu/abs/2004ApJ...613..898T}
}

@ARTICLE{Lagos2011,
       author = {{Lagos}, Claudia Del P. and {Baugh}, Carlton M. and {Lacey}, Cedric G. and {Benson}, Andrew J. and {Kim}, Han-Seek and {Power}, Chris},
        title = "{Cosmic evolution of the atomic and molecular gas contents of galaxies}",
      journal = {\mnras},
         year = 2011,
        month = dec,
       volume = {418},
       number = {3},
        pages = {1649-1667},
          doi = {10.1111/j.1365-2966.2011.19583.x},
archivePrefix = {arXiv},
       eprint = {1105.2294},
 primaryClass = {astro-ph.CO},
       adsurl = {https://ui.adsabs.harvard.edu/abs/2011MNRAS.418.1649L}
}

@ARTICLE{Bigiel2008,
       author = {{Bigiel}, F. and {Leroy}, A. and {Walter}, F. and {Brinks}, E. and {de Blok}, W.~J.~G. and {Madore}, B. and {Thornley}, M.~D.},
        title = "{The Star Formation Law in Nearby Galaxies on Sub-Kpc Scales}",
      journal = {\aj},
         year = 2008,
        month = dec,
       volume = {136},
       number = {6},
        pages = {2846-2871},
          doi = {10.1088/0004-6256/136/6/2846},
archivePrefix = {arXiv},
       eprint = {0810.2541},
 primaryClass = {astro-ph},
       adsurl = {https://ui.adsabs.harvard.edu/abs/2008AJ....136.2846B}
}

@ARTICLE{Kennicutt2012,
       author = {{Kennicutt}, Robert C. and {Evans}, Neal J.},
        title = "{Star Formation in the Milky Way and Nearby Galaxies}",
      journal = {\araa},
         year = 2012,
        month = sep,
       volume = {50},
        pages = {531-608},
          doi = {10.1146/annurev-astro-081811-125610},
archivePrefix = {arXiv},
       eprint = {1204.3552},
 primaryClass = {astro-ph.GA},
       adsurl = {https://ui.adsabs.harvard.edu/abs/2012ARA&A..50..531K}
}

@ARTICLE{Hou2000,
       author = {{Hou}, J.~L. and {Prantzos}, N. and {Boissier}, S.},
        title = "{Abundance gradients and their evolution in the Milky Way disk}",
      journal = {\aap},
         year = 2000,
        month = oct,
       volume = {362},
        pages = {921-936},
          doi = {10.48550/arXiv.astro-ph/0007164},
archivePrefix = {arXiv},
       eprint = {astro-ph/0007164},
 primaryClass = {astro-ph},
       adsurl = {https://ui.adsabs.harvard.edu/abs/2000A&A...362..921H}
}

@ARTICLE{MM2011,
       author = {{Mu{\~n}oz-Mateos}, J.~C. and {Boissier}, S. and {Gil de Paz}, A. and {Zamorano}, J. and {Kennicutt}, Jr., R.~C. and {Moustakas}, J. and {Prantzos}, N. and {Gallego}, J.},
        title = "{Radial Distribution of Stars, Gas, and Dust in SINGS Galaxies. III. Modeling the Evolution of the Stellar Component in Galaxy Disks}",
      journal = {\apj},
         year = 2011,
        month = apr,
       volume = {731},
       number = {1},
          eid = {10},
        pages = {10},
          doi = {10.1088/0004-637X/731/1/10},
archivePrefix = {arXiv},
       eprint = {1102.1724},
 primaryClass = {astro-ph.CO},
       adsurl = {https://ui.adsabs.harvard.edu/abs/2011ApJ...731...10M}
}

@ARTICLE{Recchi2008,
       author = {{Recchi}, S. and {Spitoni}, E. and {Matteucci}, F. and {Lanfranchi}, G.~A.},
        title = "{The effect of differential galactic winds on the chemical evolution of galaxies}",
      journal = {\aap},
         year = 2008,
        month = oct,
       volume = {489},
       number = {2},
        pages = {555-565},
          doi = {10.1051/0004-6361:200809879},
archivePrefix = {arXiv},
       eprint = {0808.0118},
 primaryClass = {astro-ph},
       adsurl = {https://ui.adsabs.harvard.edu/abs/2008A&A...489..555R}
}

@ARTICLE{Schonrich2009,
       author = {{Sch{\"o}nrich}, Ralph and {Binney}, James},
        title = "{Chemical evolution with radial mixing}",
      journal = {\mnras},
         year = 2009,
        month = jun,
       volume = {396},
       number = {1},
        pages = {203-222},
          doi = {10.1111/j.1365-2966.2009.14750.x},
archivePrefix = {arXiv},
       eprint = {0809.3006},
 primaryClass = {astro-ph},
       adsurl = {https://ui.adsabs.harvard.edu/abs/2009MNRAS.396..203S}
}

@ARTICLE{Spitoni2019,
       author = {{Spitoni}, E. and {Cescutti}, G. and {Minchev}, I. and {Matteucci}, F. and {Silva Aguirre}, V. and {Martig}, M. and {Bono}, G. and {Chiappini}, C.},
        title = "{2D chemical evolution model: The impact of Galactic disc asymmetries on azimuthal chemical abundance variations}",
      journal = {\aap},
         year = 2019,
        month = aug,
       volume = {628},
          eid = {A38},
        pages = {A38},
          doi = {10.1051/0004-6361/201834665},
archivePrefix = {arXiv},
       eprint = {1811.11196},
 primaryClass = {astro-ph.GA},
       adsurl = {https://ui.adsabs.harvard.edu/abs/2019A&A...628A..38S}
}

@ARTICLE{Ferguson2001,
       author = {{Ferguson}, A.~M.~N. and {Clarke}, C.~J.},
        title = "{The evolution of stellar exponential discs}",
      journal = {\mnras},
         year = 2001,
        month = aug,
       volume = {325},
       number = {2},
        pages = {781-791},
          doi = {10.1046/j.1365-8711.2001.04501.x},
archivePrefix = {arXiv},
       eprint = {astro-ph/0103205},
 primaryClass = {astro-ph},
       adsurl = {https://ui.adsabs.harvard.edu/abs/2001MNRAS.325..781F}
}

@ARTICLE{Sellwood1993,
       author = {{Sellwood}, J.~A. and {Wilkinson}, A.},
        title = "{Dynamics of barred galaxies}",
      journal = {Reports on Progress in Physics},
         year = 1993,
        month = feb,
       volume = {56},
       number = {2},
        pages = {173-256},
          doi = {10.1088/0034-4885/56/2/001},
archivePrefix = {arXiv},
       eprint = {astro-ph/0608665},
 primaryClass = {astro-ph},
       adsurl = {https://ui.adsabs.harvard.edu/abs/1993RPPh...56..173S}
}

@ARTICLE{Rovskar2008,
       author = {{Ro{\v{s}}kar}, Rok and {Debattista}, Victor P. and {Stinson}, Gregory S. and {Quinn}, Thomas R. and {Kaufmann}, Tobias and {Wadsley}, James},
        title = "{Beyond Inside-Out Growth: Formation and Evolution of Disk Outskirts}",
      journal = {\apjl},
         year = 2008,
        month = mar,
       volume = {675},
       number = {2},
        pages = {L65},
          doi = {10.1086/586734},
archivePrefix = {arXiv},
       eprint = {0710.5523},
 primaryClass = {astro-ph},
       adsurl = {https://ui.adsabs.harvard.edu/abs/2008ApJ...675L..65R}
}

@ARTICLE{Tinsley1980,
       author = {{Tinsley}, B.~M.},
        title = "{Evolution of the Stars and Gas in Galaxies}",
      journal = {\fcp},
         year = 1980,
        month = jan,
       volume = {5},
        pages = {287-388},
          doi = {10.48550/arXiv.2203.02041},
archivePrefix = {arXiv},
       eprint = {2203.02041},
 primaryClass = {astro-ph.GA},
       adsurl = {https://ui.adsabs.harvard.edu/abs/1980FCPh....5..287T}
}

@ARTICLE{Lee2024,
       author = {{Lee}, Amanda M. and {Koda}, Jin and {Hirota}, Akihiko and {Egusa}, Fumi and {Heyer}, Mark},
        title = "{The CO-to-H$_{2}$ Conversion Factor in the Barred Spiral Galaxy M83}",
      journal = {\apj},
         year = 2024,
        month = jun,
       volume = {968},
       number = {2},
          eid = {97},
        pages = {97},
          doi = {10.3847/1538-4357/ad40a0},
archivePrefix = {arXiv},
       eprint = {2404.14503},
 primaryClass = {astro-ph.GA},
       adsurl = {https://ui.adsabs.harvard.edu/abs/2024ApJ...968...97L}
}

@ARTICLE{Sextl2023,
       author = {{Sextl}, Eva and {Kudritzki}, Rolf-Peter and {Zahid}, H. Jabran and {Ho}, I.-Ting},
        title = "{Mass-Metallicity Relationship of SDSS Star-forming Galaxies: Population Synthesis Analysis and Effects of Star Burst Length, Extinction Law, Initial Mass Function, and Star Formation Rate}",
      journal = {\apj},
         year = 2023,
        month = jun,
       volume = {949},
       number = {2},
          eid = {60},
        pages = {60},
          doi = {10.3847/1538-4357/acc579},
archivePrefix = {arXiv},
       eprint = {2303.11024},
 primaryClass = {astro-ph.GA},
       adsurl = {https://ui.adsabs.harvard.edu/abs/2023ApJ...949...60S}
}

@ARTICLE{Catinella2018,
       author = {{Catinella}, Barbara and {Saintonge}, Am{\'e}lie and {Janowiecki}, Steven and {Cortese}, Luca and {Dav{\'e}}, Romeel and {Lemonias}, Jenna J. and {Cooper}, Andrew P. and {Schiminovich}, David and {Hummels}, Cameron B. and {Fabello}, Silvia and {Ger{\'e}b}, Katinka and {Kilborn}, Virginia and {Wang}, Jing},
        title = "{xGASS: total cold gas scaling relations and molecular-to-atomic gas ratios of galaxies in the local Universe}",
      journal = {\mnras},
         year = 2018,
        month = may,
       volume = {476},
       number = {1},
        pages = {875-895},
          doi = {10.1093/mnras/sty089},
archivePrefix = {arXiv},
       eprint = {1802.02373},
 primaryClass = {astro-ph.GA},
       adsurl = {https://ui.adsabs.harvard.edu/abs/2018MNRAS.476..875C}
}

@ARTICLE{Phillipps1991,
       author = {{Phillipps}, S. and {Edmunds}, M.~G.},
        title = "{On the cause of metallicity gradients in spiral galaxies.}",
      journal = {\mnras},
         year = 1991,
        month = jul,
       volume = {251},
        pages = {84},
          doi = {10.1093/mnras/251.1.84},
       adsurl = {https://ui.adsabs.harvard.edu/abs/1991MNRAS.251...84P}
}

@ARTICLE{Werk2011,
       author = {{Werk}, J.~K. and {Putman}, M.~E. and {Meurer}, G.~R. and {Santiago-Figueroa}, N.},
        title = "{Metal Transport to the Gaseous Outskirts of Galaxies}",
      journal = {\apj},
         year = 2011,
        month = jul,
       volume = {735},
       number = {2},
          eid = {71},
        pages = {71},
          doi = {10.1088/0004-637X/735/2/71},
archivePrefix = {arXiv},
       eprint = {1104.3897},
 primaryClass = {astro-ph.CO},
       adsurl = {https://ui.adsabs.harvard.edu/abs/2011ApJ...735...71W}
}

@ARTICLE{Kewley2026,
       author = {{Kewley}, Lisa J. and {Grasha}, Kathryn and {Garcia}, Alex and {Torrey}, Paul and {Rich}, Jeff and {Hemler}, Z.~S. and {Chen}, Qian-Hui and {Zhu}, Peixin and {Seibert}, Mark and {Hernquist}, Lars and {Madore}, Barry},
        title = "{The assembly history of NGC 1365 through chemical archaeology}",
      journal = {Nature Astronomy},
         year = 2026,
        month = mar,
          doi = {10.1038/s41550-026-02808-7},
       adsurl = {https://ui.adsabs.harvard.edu/abs/2026NatAs.tmp...62K}
}

@ARTICLE{Ogami2026,
       author = {{Ogami}, Itsuki and {Okamoto}, Sakurako and {Ferguson}, Annette M.~N. and {Komiyama}, Yutaka and {Chiba}, Masashi and {Koda}, Jin and {Hayashi}, Kohei and {Suzuki}, Yoshihisa},
        title = "{A Long Stellar Stream in M83: Possible Connection Between XUV Disks and Minor Mergers?}",
      journal = {arXiv e-prints},
         year = 2026,
        month = mar,
          eid = {arXiv:2603.23981},
        pages = {arXiv:2603.23981},
          doi = {10.48550/arXiv.2603.23981},
archivePrefix = {arXiv},
       eprint = {2603.23981},
 primaryClass = {astro-ph.GA},
       adsurl = {https://ui.adsabs.harvard.edu/abs/2026arXiv260323981O}
}

@ARTICLE{Malin1997,
       author = {{Malin}, David and {Hadley}, Brian},
        title = "{HI in Shell Galaxies and Other Merger Remnants}",
      journal = {\pasa},
         year = 1997,
        month = apr,
       volume = {14},
       number = {1},
        pages = {52-58},
          doi = {10.1071/AS97052},
       adsurl = {https://ui.adsabs.harvard.edu/abs/1997PASA...14...52M}
}

@ARTICLE{Blair2014,
       author = {{Blair}, William P. and {Chandar}, Rupali and {Dopita}, Michael A. and {Ghavamian}, Parviz and {Hammer}, Derek and {Kuntz}, K.~D. and {Long}, Knox S. and {Soria}, Roberto and {Whitmore}, Bradley C. and {Winkler}, P. Frank},
        title = "{An Expanded HST/WFC3 Survey of M83: Project Overview and Targeted Supernova Remnant Search}",
      journal = {\apj},
         year = 2014,
        month = jun,
       volume = {788},
       number = {1},
          eid = {55},
        pages = {55},
          doi = {10.1088/0004-637X/788/1/55},
archivePrefix = {arXiv},
       eprint = {1404.7179},
 primaryClass = {astro-ph.GA},
       adsurl = {https://ui.adsabs.harvard.edu/abs/2014ApJ...788...55B}
}

@ARTICLE{Thilker2005,
       author = {{Thilker}, David A. and {Bianchi}, Luciana and {Boissier}, Samuel and {Gil de Paz}, Armando and {Madore}, Barry F. and {Martin}, D. Christopher and {Meurer}, Gerhardt R. and {Neff}, Susan G. and {Rich}, R. Michael and {Schiminovich}, David and {Seibert}, Mark and {Wyder}, Ted K. and {Barlow}, Tom A. and {Byun}, Yong-Ik and {Donas}, Jose and {Forster}, Karl and {Friedman}, Peter G. and {Heckman}, Timothy M. and {Jelinsky}, Patrick N. and {Lee}, Young-Wook and {Malina}, Roger F. and {Milliard}, Bruno and {Morrissey}, Patrick and {Siegmund}, Oswald H.~W. and {Small}, Todd and {Szalay}, Alex S. and {Welsh}, Barry Y.},
        title = "{Recent Star Formation in the Extreme Outer Disk of M83}",
      journal = {\apjl},
         year = 2005,
        month = jan,
       volume = {619},
       number = {1},
        pages = {L79-L82},
          doi = {10.1086/425251},
archivePrefix = {arXiv},
       eprint = {astro-ph/0411306},
 primaryClass = {astro-ph},
       adsurl = {https://ui.adsabs.harvard.edu/abs/2005ApJ...619L..79T}
}

@ARTICLE{Sextl2026,
       author = {{Sextl}, Eva and {Kudritzki}, Rolf-Peter},
        title = "{The Hidden Story of Chemical Evolution in Local Star-forming Nuclear Rings}",
      journal = {\apj},
         year = 2026,
        month = feb,
       volume = {997},
       number = {2},
          eid = {329},
        pages = {329},
          doi = {10.3847/1538-4357/ae29ee},
archivePrefix = {arXiv},
       eprint = {2510.14757},
 primaryClass = {astro-ph.GA},
       adsurl = {https://ui.adsabs.harvard.edu/abs/2026ApJ...997..329S}
}

@ARTICLE{Long2022,
       author = {{Long}, Knox S. and {Blair}, William P. and {Winkler}, P. Frank and {Della Bruna}, Lorenza and {Adamo}, Angela and {McLeod}, Anna F. and {Amram}, Phillippe},
        title = "{Supernova Remnants in M83 as Observed with MUSE}",
      journal = {\apj},
         year = 2022,
        month = apr,
       volume = {929},
       number = {2},
          eid = {144},
        pages = {144},
          doi = {10.3847/1538-4357/ac5aa3},
archivePrefix = {arXiv},
       eprint = {2202.09929},
 primaryClass = {astro-ph.HE},
       adsurl = {https://ui.adsabs.harvard.edu/abs/2022ApJ...929..144L}
}

@ARTICLE{Molla2017,
       author = {{Moll{\'a}}, Mercedes and {D{\'\i}az}, {\'A}ngeles I. and {Ascasibar}, Yago and {Gibson}, Brad K.},
        title = "{Galaxy chemical evolution models: the role of molecular gas formation}",
      journal = {\mnras},
         year = 2017,
        month = jun,
       volume = {468},
       number = {1},
        pages = {305-318},
          doi = {10.1093/mnras/stx419},
archivePrefix = {arXiv},
       eprint = {1702.04334},
 primaryClass = {astro-ph.GA},
       adsurl = {https://ui.adsabs.harvard.edu/abs/2017MNRAS.468..305M}
}

\end{document}